\documentclass[prl,superscriptaddress,preprint,showpacs,letterpaper,10pt,twocolumn]{revtex4}
\usepackage{color,graphicx}
\usepackage{bm}
\usepackage{amsmath}
\usepackage{amssymb}
\usepackage{epstopdf}
\usepackage{ulem}
\usepackage{url}

\def\be{\begin{equation}}
\def\ee{\end{equation}}
\def\bea{\begin{eqnarray}}
\def\eea{\end{eqnarray}}

\newcommand\mc            {\mathcal}

\renewcommand\th          {\theta}

\begin{document}

\title{Finite temperature spin dynamics in a perturbed quantum critical Ising chain with an $E_8$ symmetry}
\author{Jianda Wu}
\affiliation{Department of Physics \& Astronomy, Rice University, Houston, Texas 77005, USA}
\author{M\'{a}rton Kormos}
\affiliation{Department of Physics \& Astronomy, Rice University, Houston, Texas 77005, USA}
\affiliation{MTA-BME ``Momentum'' Statistical Field Theory Research Group, 1111 Budapest, Budafoki \'ut 8, Hungary}
\author{Qimiao Si}
\affiliation{Department of Physics \& Astronomy, Rice University, Houston, Texas 77005, USA}

\begin{abstract}
A spectrum exhibiting $E_8$ symmetry is expected to arise when a small longitudinal field is introduced in the
transverse-field Ising chain at its quantum critical point.
Evidence for this spectrum has recently
come from neutron scattering measurements in cobalt niobate, a quasi-one-dimensional Ising  ferromagnet.
Unlike its zero-temperature counterpart, the
finite-temperature dynamics of the model
has not yet been determined. We study the dynamical spin structure factor of the model at low frequencies and nonzero
temperatures,
using the form factor method. Its frequency dependence
is singular,
but differs from the diffusion form. The temperature
dependence of the nuclear magnetic resonance (NMR) relaxation rate has an activated form,
whose prefactor we also determine. We propose NMR experiments as a means
to further test the applicability of the $E_8$ description for CoNb$_2$O$_6$.
\end{abstract}
\pacs{71.10.Hf,  73.43.Nq, 71.10.-w.}
\maketitle

\paragraph*{Introduction.---}
Quantum criticality is a subject of extensive interest in various contexts \cite{JLTP-issue10,sachdev}.
These range from
correlated-electron bulk materials, which can be tuned to the border of magnetism,
to systems in low dimensions, where quantum fluctuations are enhanced.
The collective fluctuations of a quantum critical point (QCP) often lead to unusual properties.
Even in equilibrium, the statics and dynamics are mixed at a QCP. This gives rise
to dynamical scaling, while also making it difficult to calculate the fluctuation spectrum.
The latter is especially so for the dynamics at nonzero temperatures ($T>0$)
in the ``quantum relaxational" regime, which corresponds to small frequencies
($\omega \ll k_B T/\hbar$) or long times. Indeed, even for the canonical QCP
of a transverse-field Ising model in one dimension, it has been challenging to calculate
such real-frequency dynamics \cite{deift,sachdev2}.

We are interested here in the one dimensional transverse field Ising model in the presence
of a small longitudinal field. The transverse-field-induced QCP in the absence of a longitudinal
field \cite{pfeuty} has an emergent conformal invariance in the scaling limit \cite{belavin}.
When a small
longitudinal field is turned on at the QCP, the excitation spectrum becomes discrete at low energies.
The perturbed
conformal field theory \cite{zamolodchikov} provided evidence that
certain properties of the spectrum of the resulting relativistic field theory and the scattering matrix
can be organized in terms of $E_8$, an exceptional simple Lie group of rank $8$.
The discrete spectrum corresponds to eight particles,
whose  masses form ratios which are related to the roots of the $E_8$ algebra.
(For introductory discussions, see Refs.~\cite{dorey,borthwick}.)
The first two particles
describe
bound states that are well below the continuum part of the spectrum.
Recently neutron scattering
measurements have been carried out in a
ferromagnetic cobalt niobate CoNb$_2$O$_6$,
whose Co$^{2+}$ are coupled in a quasi-1D way; the experiment
identified two excitations whose energy ratios
are close to the predicted value, the golden ratio \cite{coldea}.

In this letter, we study the low-frequency dynamical spin structure factor at finite temperatures
using the form factor method \cite{mussardo}.
From a theoretical perspective, our calculation
provides an illustrative
setting to determine the dynamics in the
quantum-relaxational regime. For the $E_8$ model, the dynamics at finite temperatures have not been systematically studied.
From the perspective of the material CoNb$_2$O$_6$, our study determines the
temperature dependence of the NMR relaxation rate.
We note that our results bear some similarities with those for another model,
the O(3) non-linear sigma model \cite{SagiAffleck,Konik03}, although our study here benefits from
the exactly regularized form
factor series \cite{pozsgay,szecsenyi}.
We also note that a numerical analysis of a generalized transverse-field Ising chain suggests that the
E$_8$ description survives suitable generalizations of the
interactions beyond the nearest-neighbor ferromagnetic
coupling \cite{kjall}.

\paragraph*{The Model.---}
Consider
the Hamiltonian
\be
H_{Z}  =  - J\left( {\sum\limits_i {\sigma _i^z \sigma _{i + 1}^z }  + g \sum\limits_i {\sigma _i^x }
+ h_z\sum\limits_i {\sigma _i^z } } \right) ,
\label{ZE8Hamiltonian}
\ee
where $\sigma _i^x$ and $\sigma _i^z$ are the Pauli matrices associated
with the spin components $S^\mu = \sigma^\mu/2, ({\mu=x,y,z})$, and $i$ marks a site position,
in addition $g$ and $h_z$ are the physical transverse and longitudinal fields,
respectively,
in unit of the nearest-neighbor ferromagnetic exchange coupling $J$ between the
longitudinal ($z$) components of the spins. In the absence of the longitudinal field ($h_z=0$)
the system undergoes a quantum phase transition when the transverse field is tuned across
its critical value $g=g_c=1$ \cite{pfeuty}. As is well known, the QCP is described by
a $1+1$-dimensional conformal field theory (CFT) with a central charge $1/2$ \cite{belavin}.
More surprising is what happens when a small longitudinal field $h_z$ is introduced at the QCP $g=g_c$.
The action in the continuum limit is given by:
\be
\mc{A}_{E8}  = \mc{A}_{c = 1/2}  + h\int dx d\tau
 \sigma (\tau,x) .
 \label{ZE8action}
\ee
This is an integrable field theory,
and is referred to as the $E_8$ model
because of the aforementioned connection between its properties and the $E_8$ group \cite{zamolodchikov,mussardo}.
In the above equation, $\tau$ is the imaginary time,
$\mc{A}_{c = 1/2}$ stands for the action of the two dimensional CFT with central charge $1/2$,
and $\sigma(x)$ is a primary field with scaling dimension $1/8$.
In addition, $h=c J h_z /a$, where $a$ is the lattice constant and
$c \approx 0.783$ converts between
the $\sigma$ field of the continuum theory and its lattice counterpart $\sigma ^z$ \cite{delfino1}.
This describes a scattering theory of eight massive particles, which we will denote by $a,b,c,d,e,f,g,h$ from the lightest to the heaviest.
The mass of the lightest particle, $\Delta_a$, scales with the longitudinal field as $\Delta _a  \approx 4.405\left| h \right|^{8/15}$ \cite{mass}.
The mass of the
second lightest particle $\Delta_b$ is $\Delta_a$ multiplied by the golden ratio $(\sqrt{5}+1)/2$.
These two particles are clearly separated from the two-particle continuum, which appears
at energies above $2\Delta_a$.

\paragraph*{Local dynamics and NMR relaxation rate.---}
We focus on the local dynamical structure factor (DSF) of the $E_8$ model in the low frequency and
low temperature limit: $\omega \ll \Delta_a$ and $T \ll \Delta_a$ (hereafter we
set $\hbar=1$ and $k_B=1$).

A useful means to probe the local DSF is via NMR.
The NMR relaxation rate
is given by \cite{moriya}
\be
\frac{1}{{T_1^{\alpha} }} = \frac{1}{{2N}}A^2 \sum_{\beta}
{}^{^{\prime}} S_{\beta\beta}(\omega _0 ) . \label{NMR}
\ee
Here, $\alpha$ and $\beta$ label the principal axes, and
the primed summation is over the
principal axes perpendicular to the field orientation $\alpha$; $T_1^{\alpha}$
is the spin-lattice relaxation time, $N$ is the number of ions per unit cell,
and $\omega_0$ is the nuclear resonance frequency.
In addition, $A$ describes the hyperfine coupling between the spins of
a nucleus and the electrons; while this coupling depends on the wavevector ${\bf q}$,
the dependence is generically smooth and we will take it as a constant.
We will consider the static field of the NMR setup to be the transverse field,
$\alpha=x$. Correspondingly, the local DSF of interest to NMR
is given by $S_{zz}(\omega _0 ) +S_{yy}(\omega _0 ).$
As shown in the supplementary materials \cite{supplement},
for the model we consider,
\be
S_{yy} (\omega) =  \omega^2 S_{zz} (\omega)/(4J^2).
\ee
Thus, in the low-frequency regime of interest here,
$S_{yy} (\omega)$ is negligible compared with $S_{zz} (\omega)$.
In the following, we will therefore only consider $S_{zz}$.

We now turn to the calculation of
$S_{zz}(\omega)$
through a systematic form factor expansion.
Because the excitation spectrum has a gap,
 we expect that the leading contributions
  in the low temperature
and low frequency limit come from
those associated with the
few particle states of the light particles.
Indeed, we show below that the dominant contribution
comes from
the two 1-particle states of the lightest particle, which we calculate analytically.
The conclusion is confirmed by a numerical calculation for contributions that extend to higher
orders.

\paragraph{The form factor series.---}
Integrable field theory techniques made possible the analytic calculation of matrix elements
of local observables in the asymptotic scattering state basis, called {\it{form factors}}.
The asymptotic states are eigenstates of the energy and momentum operators.
It is convenient to use the standard reparameterization in relativistic theories of a particle's energy and momentum
through the rapidity of the particle.
In terms of the rapidities $\{\theta_i\}$ of the particles, the energy and momentum eigenvalues
of the eigenstate $\left| {\theta _1^{\alpha_1} , \cdots ,\theta _n^{\alpha_n} } \right\rangle$ (with
$\{\alpha_1, \cdots, \alpha_n \}$ marking different types of particles)
are
\begin{align}
E_n &= \sum_{i=1}^n \Delta_{\alpha_i}\cosh(\theta_i),\\
P_n &= \sum_{i=1}^n \Delta_{\alpha_i}\sinh(\theta_i).
\end{align}
We denote by $F_n^{\sigma}(\theta_1^{\alpha_1},\cdots,\theta_n^{\alpha_n})$ the form factors
of the primary field $\sigma(t, x)$ in the $E_8$ model (c.f. Eq.~\eqref{ZE8action})
between the vacuum and an $n$-particle asymptotic state,
\be
F_n^\sigma  (\theta _1^{\alpha_1} , \cdots ,\theta _n^{\alpha_n} ) = \left\langle 0 \right|\sigma(0,0)
\left| {\theta _1^{\alpha_1} , \cdots ,\theta _n^{\alpha_n} } \right\rangle. \label{formfactor}
\ee
The few-particle form factors are explicitly known \cite{delfino1,delfino2, delfino3} and have been used to calculate
the static spin-spin correlations of the
$E_8$ model in the ground state  \cite{delfino1,delfino2}.
Here we study the finite-temperature dynamics by a low-temperature expansion series for integrable field theory \cite{essler,pozsgay}, using a finite-volume regularization
\cite{pozsgay}.

The finite temperature two-point correlation function is given by
\be
C(t,x)=\mathrm{Tr}\left[\frac{e^{-H/T}}{\mc{Z}} \mc{O}(t,x)\mc{O}^\dagger(0,0)\right], \label{correlation1}
\ee
where $\mc{Z}=\mathrm{Tr}\,e^{-H/T}$ is the partition function,
and we are interested in the local observable operator $\mc{O}(t,x) = \sigma(t,x)$ .
The corresponding DSF is
\be
S(\omega ,q) = \int_{ - \infty }^\infty  {dx\int_{ - \infty }^\infty  {dt\;C(t,x)e^{i\omega t - iqx} } },
\label{dynamicstructurefactor}
\ee
We insert the complete set of asymptotic states between the operators, yielding
a double sum, $C(t,x)=\mathcal{Z}^{-1}\sum_{r,s} C_{r,s}(t,x)$, where
\begin{multline}
C_{r,s}(t,x) = \sum\limits_{\left\{ {\alpha _j } \right\},\left\{ {\alpha '_k } \right\}} {\int {\frac{{d\theta _1  \cdots d\theta _r }}{{(2\pi )^r r!}}\int {\frac{{d\theta '_1  \cdots d\theta '_s }}{{(2\pi )^s s!}}} } }e^{ - \beta E_r } \\
  e^{ - it(E_s  - E_r )} e^{ - i(P_r  - P_s )x} \left| {\left\langle {\theta _1^{\alpha _1 }  \cdots \theta _r^{\alpha _r } } \right|\mc{O}\left| {\theta '{_1^{\alpha '_1 }}  \cdots \theta'{_s^{\alpha '_s } }} \right\rangle }  \right|^2.
\end{multline}
%

We use the same set of states to write the partition function as
$\mc{Z}=\sum_{n=0}^\infty \mc{Z}_n$ where
\be
\mc{Z}_n = \sum_{\{ \alpha _j \}} \int \frac{d\theta _1  \cdots d\theta _n }{(2\pi )^n n!} e^{-\beta E_n}
\left\langle \theta _1^{\alpha _1 }  \cdots \theta _n^{\alpha _n }  \right| \theta_1^{\alpha _1 }  \cdots \theta_n^{\alpha_n } \rangle .
\ee
In infinite volume all the $\mc{Z}_n$'s contain
singularities associated with
the scalar product of two
momentum eigenstates with identical rapidities.
Similarly, for the observables we are calculating, $C_{r,s}$
also diverge due to
the kinematical poles of the form factors whenever two rapidities in
the two sets coincide, $\th_i=\th'_j$ \cite{essler}. However,
the double sums can be re-organized such that the aforementioned singularities cancel
each other \cite{pozsgay},
\be
C(t,x)=\sum_{r,s=0}^\infty D_{r,s}(t,x)\,, \label{correlationseries}
\ee
where
\begin{align}
D_{0,s}&=C_{0,s}\,, \label{d0s}\\
D_{1,s}&=C_{1,s}-\mc{Z}_1C_{0,s-1}\,,\label{d1s} \\
D_{2,s}&=C_{2,s}-\mc{Z}_1C_{1,s-1}+(\mc{Z}_1^2-\mc{Z}_2)C_{0,s-2},\\ \label{d2s}
\text{\dots etc.}
\end{align}

The natural small parameter in the series \eqref{correlationseries} is $e^{-\Delta_a/T}$.
At low frequencies, the energy conserving Dirac-deltas in the Fourier transform Eq. \eqref{dynamicstructurefactor}
force the two states appearing
in the form factors to have nearly equal energy, $E_r=\omega+E_s$.
The magnitude of the Boltzmann factor is then set by the sum of the masses in the ``heavier" state,
i.e.,
\be
D_{r,s}  \sim \exp \left\{ { - \frac{1}{T}\max \left[ {\sum\limits_{i = 1}^r {\Delta _i } ,
\sum\limits_{i = 1}^s {\Delta _i } } \right]} \right\}.
 \label{leadingordercounting}
 \ee
Thus, in the regime of interest ($T/\Delta_a \ll 1$ and $\omega/\Delta_a \ll 1$),
the expansion series in Eq.~(\ref{correlationseries}) is a good
perturbation series.
In this regime, we can safely truncate the series beyond the terms up to the
order of $e^{-2 \Delta_a/T}$. Simple counting implies that we only need
 $D_{0,1}, D_{1,0}, D_{0,2}, D_{2,0}, D_{1,1},D_{1,2}+D_{2,1},D_{2,2} $ with lightest particles,
which we now determine.
We also note that the series for the two-point correlator {\it per se} contain a $\delta(\omega)$ piece, which are however absent
in the connected correlation function of interest here \cite{supplement}.

\paragraph{Leading contributions.---}

$D_{0,1}$ is the channel between vacuum and one-particle asymptotic
``in'' state, and is equal to $C_{0,1}$ from Eq.~(\ref{d0s}). The corresponding contribution
to DSF is
\be S_{0,1}(\omega,q) = 2\pi \left| {F_1^\sigma  }
\right|^2 \int {d\theta \delta (q - \Delta _1 \sinh \theta )\delta
\left( {\omega  - \Delta _1 \cosh \theta } \right)},
\label{dynamic01}
\ee
where $\Delta_1$ is the mass of a single
particle state, and the one particle form factor $F_1^\sigma (\theta)$
is rapidity independent\cite{delfino1}. Since $\cosh \theta \geq 1$ always holds, for
the parameter regime $\omega<\Delta_a $ the terms $S_{0,1}$ and $S_{1,0}$
do not contribute. Similarly, the $D_{0,s}$ and $D_{r,0}$ terms for general $r$ and $s$
also vanish.

The first non-trivial contribution is given by connected parts in $D_{1,1}$, i.e. the term coming from
the 1-particle -- 1-particle form factors, for which we obtain \cite{supplement}
\be
S_{1,1} (\omega ,q) = \frac{{\left| {F_2^\sigma  (\alpha + i\pi, 0 )} \right|^2
\left(e^{ - \beta \Delta _1 \cosh \theta _{+} } +e^{ - \beta \Delta _1 \cosh
\theta _{-} } \right) }}{{\Delta _1 \Delta _2 |\sinh  \alpha| }} ,
\label{S11gen}
\ee
where $\Delta _1$ and $\Delta _2$ are the masses of the 1-particle states,
$\alpha  = {\rm{arccosh}}[(\Delta _1^2  + \Delta _2^2  - (\omega ^2  - q^2 ))/(2\Delta _1 \Delta _2 )]$
and $\cosh \theta _{\pm}  = [\omega (\Delta _1^2  - \Delta _2^2  + \omega ^2  - q^2 )
\pm 2q\Delta _1 \Delta _2 \sinh\alpha]/[2\Delta _1 (q^2  - \omega ^2 )]$; hereafter the symbols that denote the types of particles in the form factor are dropped for notational convenience [Eq.~(\ref{formfactor})].

The corresponding local DSF is $S_{1,1} (\omega ) = \int_{ - \infty }^\infty
{S_{1,1} (q,\omega )dq}$.
Eq.~(\ref{leadingordercounting}) implies that, up to $e^{-2\Delta_a/T}$, we need only to consider
the channels $a-a$, $b-b$ and $c-c$, as well as
$a-b$, $a-c$, $b-c$.
When $\Delta_1= \Delta_2=\Delta_i$ ($i=a,\dots,h$),
\be
\left. {S_{1,1} (\omega )} \right|_{\Delta _1  = \Delta _2  = \Delta _i }
= \int_\omega ^\infty  {f(q,\omega )e^{ - \frac{{\Delta _i }}{T}g(q,\omega )} dq}
\label{S11equalmass}
\ee
with $f(q,\omega ) = \frac{2}{\Delta _i^2 }\left| F_2^\sigma  (\alpha + i\pi,0 )
\right|^2 /|\sinh\alpha|$ and $g(q,\omega ) =   - \frac{\omega }{2\Delta _i }
+ \frac{q}{2\Delta _i }\sqrt {1+\frac{4\Delta _i^2 }{q^2  - \omega ^2 } }.$
We can expand the result for small $\omega$.
With the details given in the supplementary material \cite{supplement},
we find the result to leading order:
\begin{widetext}
 \bea
\left. {S_{1,1} (\omega )} \right|_{\Delta _1  = \Delta _2  = \Delta _a }
 \approx \left\{ \begin{array}{l}
  \frac{{2 \left| {F_2^\sigma \left( {i\pi, 0 } \right)}
\right|^2 }}{\Delta_a }
  e^{ - \Delta_a /T} \left\{ {\ln \frac{4T }{{\omega}} -\gamma _E
  +\cdots} \right\} \;(\omega \ll T  \ll \Delta_a ) \\
 \frac{{2 \left| {F_2^\sigma \left( {i\pi, 0 } \right)}
\right|^2}}{{\Delta _a }} {e^{ - \Delta _a /T}}
 \left\{ {\sqrt {\frac{{\pi T}}{\omega }}  - \frac{{\sqrt \pi  }}{4}\left( {\frac{T}{\omega }}
 \right)^{3/2}  +  \cdots} \right\}\;(T \ll \omega  \ll \Delta_a ) \\
 \end{array} \right.,
\label{S11leading}
 \eea
\end{widetext}
where $\gamma_E$ is the Euler constant. (The same form applies to the contributions by the
other particles $b,\cdots,h$, which are suppressed by their thermal factors.)
 In deriving this expression, we have replaced
$\alpha(\omega,q)$ by $\alpha(\omega=0,q=0)$.
This is because the dominant contribution comes
from the minimum of the energy dispersion at small momentum; it is well supported by the numerical calculation carried out without this replacement (see below).

We observe that the finite-T local DSF
diverges logarithmically as $\omega\to0$.
This divergence differs
from the diffusion form \cite{giuliani} of inverse square
root; this is reasonable
 given that the total $S_z$ is not conserved here. When $\Delta_1 \neq \Delta_2$, the denominator
 on the right hand side of Eq.~(\ref{S11gen}) does not have any singularity so there will be no divergence.

\begin{figure}[t!]
\begin{center}
\includegraphics[width=8.25cm]{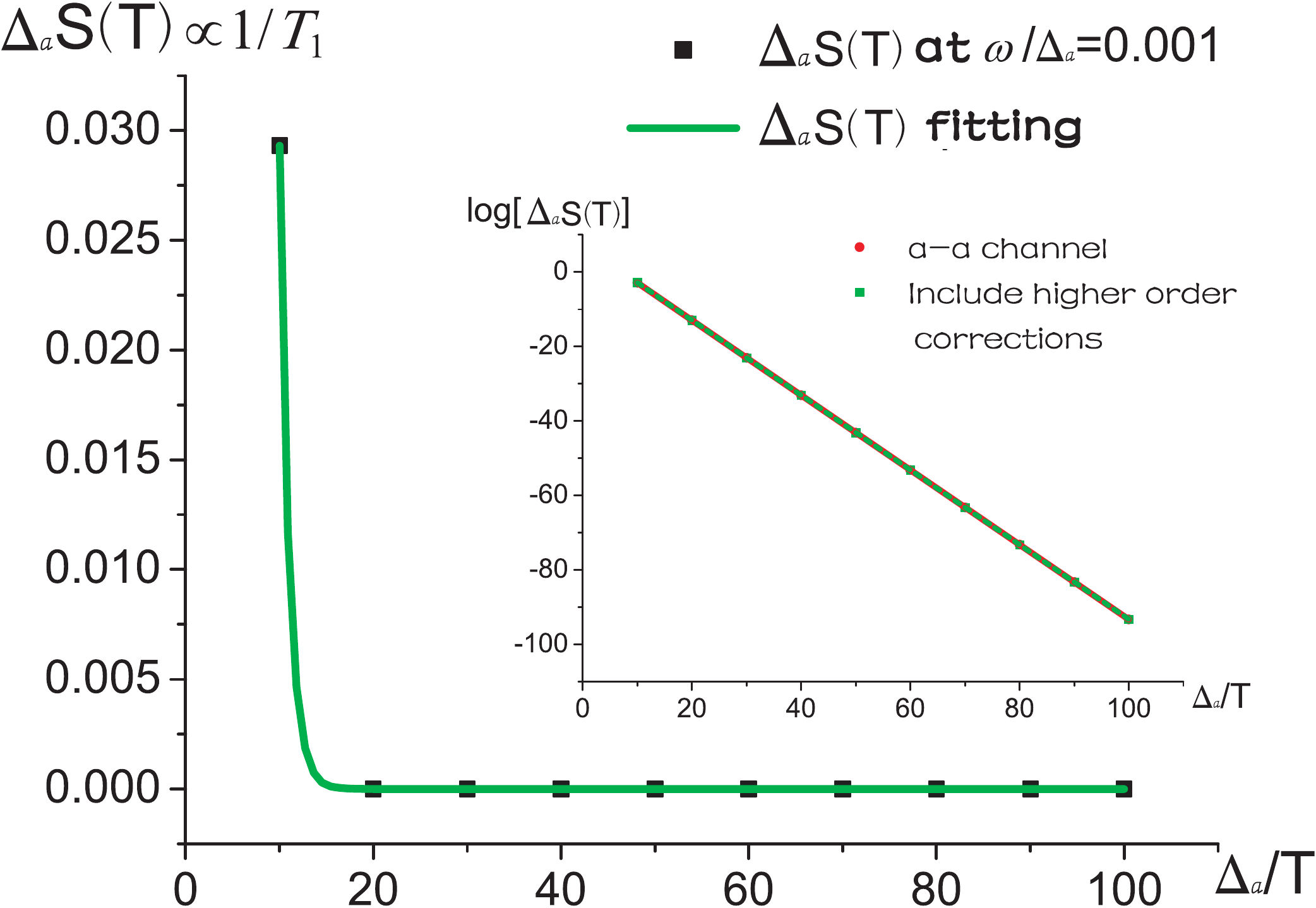}
\end{center}
\caption{The NMR relaxation rate as a function of temperature. The frequency is chosen to be
$\omega/\Delta_a = 0.001$. The temperature dependence is well described by
$\Delta_a S(T) = 631 e^{- \Delta_a/T}$.
The inset picture shows that channels other than $a-a$ give negligible contributions.}
\label{fig:relaxationrate}
\end{figure}

Next, we consider $D_{1,2}+D_{2,1}$, the terms with a one-particle and a two-particle state.
Up to the order $\mathcal{O}(e^{-2\Delta_a/T})$, we focus on the case
when all three particles are the lightest $a$ particle (the other channels $aa-b$ and $aa-c$ are expected to behave similarly), which we find to be \cite{supplement},
\begin{multline}
S_{(1,2)+(2,1)} (\omega,q)  = \frac{1}{{\pi }}\int_{ - \infty }^\infty  {\frac{{d\theta  e^{ - \beta \Delta_a\cosh \theta  }}}{{\left| {\sqrt {\left( {f\tilde\omega ,\tilde q,\theta  ) - 1} \right)^2  - 1} } \right|}} }\\
 \cdot F_3^\sigma  \left( {\theta   + i\pi ,\ln x_+,\ln x_-} \right)F_3^\sigma  \left( {\theta  + i\pi ,\ln x_-,\ln x_+} \right),
 \label{dynamic12I}
\end{multline}
where
\be
x_\pm =
\frac{1}{2}(\tilde\omega  + \cosh \theta   + \tilde q + \sinh \theta)
\left(1   \pm  2\sqrt{1-2/f(\tilde\omega,\tilde q,\theta)}\right),
\ee
and $f(\tilde\omega ,\tilde q,\theta ) = \left[ {\left( {\tilde\omega  + \cosh \theta  } \right)^2
- \left( {\tilde q + \sinh \theta  } \right)^2 } \right]/2$ with $\tilde \omega=\omega/\Delta_a$
and $\tilde q=q/\Delta_a.$
Our analysis \cite{supplement}
shows
no contributions from the range
$\tilde\omega > \tilde q \geq 0 $,
where $\cosh \theta \sim 1/\tilde\omega \gg 1$.
In the range $\tilde\omega \leq \tilde q $,
we have $\cosh\theta \gtrsim 2-\tilde\omega$, indicating there exists a small region of $\tilde q$ where $\cosh\theta$
is slightly smaller than $2$. This contribution is expected to be small, and we confirm
this by including the channels $D_{1,2}+D_{2,1}$ in our numerical calculation shown below.

For connected parts in $D_{2,2}$, a similar Jacobian will appear as in the calculation of the equal mass
case of Eq.~(\ref{S11equalmass}), and we will encounter the same logarithmic divergence
in the frequency dependence. We find no singular terms beyond the logarithmic divergence \cite{supplement}.
 This contribution is therefore suppressed by the
thermal weight $e^{-2\Delta_a/T}$.
Low-frequency divergences are also expected to come from the $D_{nn}$ terms (at $n>2$)
with particles of the same mass in the two asymptotic states of the form factors.
The fact that $D_{22}$ with the same particle
does not contain singularities stronger than $\ln\omega$  is a strong indication that
none of the higher terms in the series will give a stronger (e.g. power-law) singularity.
We conjecture that the $D_{nn}$ terms at $n>2$ have a similar logarithmic singularity in the frequency dependence,
and they are then also negligible compared to $D_{11}$ due to the stronger thermal suppression factor.

\paragraph{Numerical analysis.---}
Fig.~\ref{fig:relaxationrate} shows
the results and fit for the NMR relaxation rate as a function of  temperature in the  range $\Delta_a/T\in [10,100]$ at a fixed low frequency $\omega/\Delta_a=0.001$ appropriate
for the NMR experiments (satisfying $\omega  \ll T $). The fitting function $\Delta_a S(T) = 631 e^{- \Delta_a/T}$ indicates that the behavior of relaxation rate at low frequency and low temperature region is dominated by the contribution from the $a-a$ channel, as clearly shown in the inset to Fig.~\ref{fig:relaxationrate}. The prefactor 631 compares well with the analytical expression associated with $S_{1,1}$ of the lightest $a$-particle:
since $2|F_2^\sigma(i\pi,0)|_{\Delta_1 =\Delta_2 =\Delta_a}^2 \approx 130$.

\begin{figure}[t!]
\begin{center}
\includegraphics[width=\columnwidth]{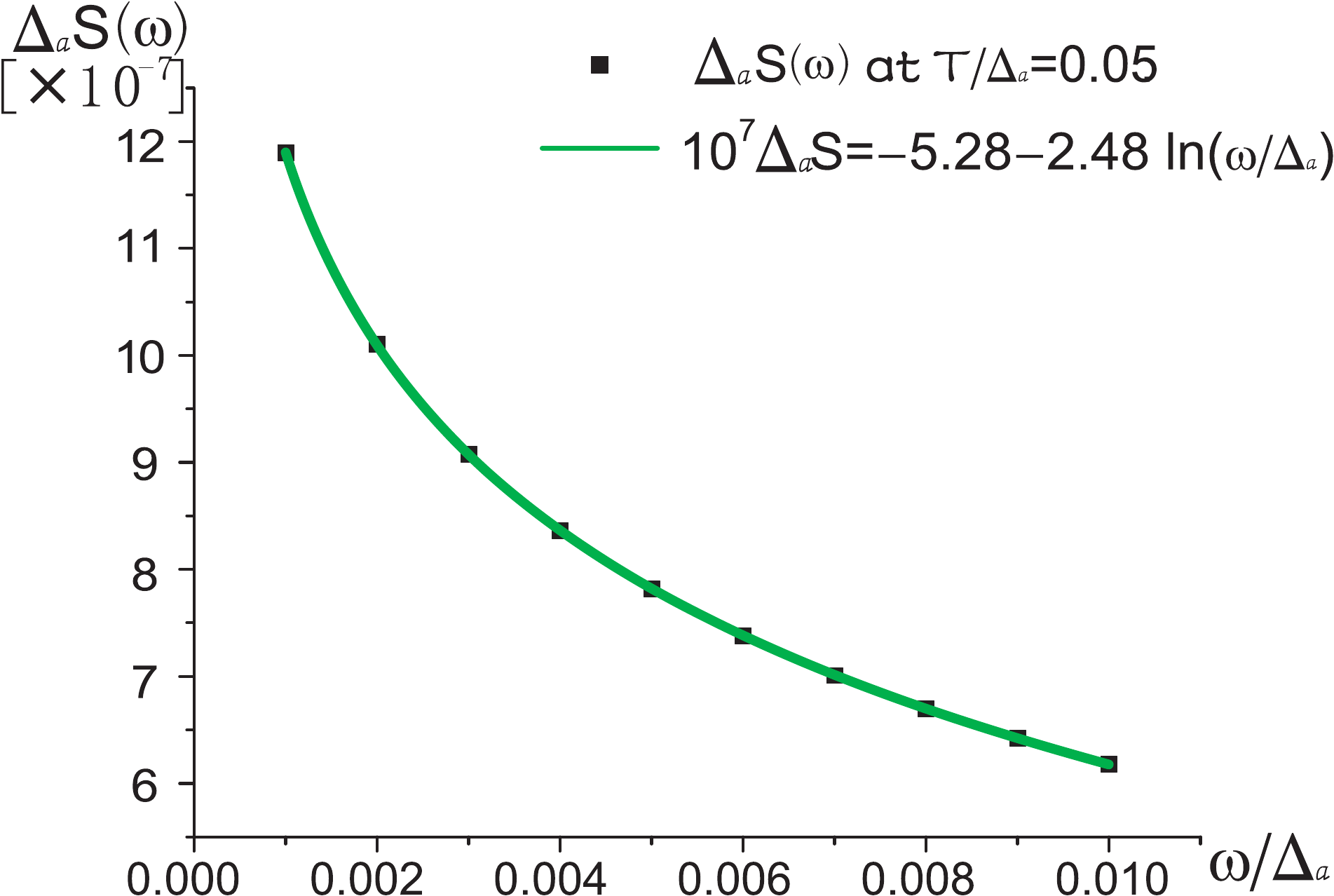}
\end{center}
\caption{The local dynamical structure factor as a function of frequency at a fixed temperature $T/\Delta_a= 0.05$.
The $\omega$-dependence is well described by $10^7 \Delta_a S(\omega) =  - 5.28 + 2.48 \ln (\Delta_a /\omega)$.  }
\label{fig:E8}
\end{figure}

We also study the frequency dependence of the local DSF at fixed temperatures
for $T,\omega \ll \Delta_a$. Fig.~\ref{fig:E8} shows the result
at a fixed $T/\Delta_a= 0.05$ with $\omega/\Delta_a$ ranging
from $0.001$ to $0.01$ (satisfying
$\omega \ll T$. It is well fitted as $10^7 \Delta_a S(\omega ) =  - 5.28 - 2.48 \ln (\omega /\Delta _a )$, which is in accordance with the asymptotic form Eq.~(\ref{S11leading}).

\paragraph*{Discussion.---}
We conclude that the temperature dependence of the NMR relaxation rate is given by
\be
\frac{1}{{T_1 }} \approx \frac{c^2 b}{{\Delta_a}} \frac{A^2}{{2N}} e^{- \Delta_a/T};\;\Delta _a  \approx 4.405\left| h \right|^{8/15}  .
\ee
In the prefactor, $c \approx 0.783$
is the aforementioned conversion factor between
the $\sigma$ field
and the lattice spin \cite{delfino1},
and $b\approx \ln(4T/\omega_0) - \gamma_E$.

We next consider the implications of our results for
CoNb$_2$O$_6$.
The neutron scattering experiments provided evidence for the two lightest particles
of the $E_8$ spectrum \cite{coldea}.
This has been understood by considering the effect of the inter-chain coupling
in the three-dimensionally ordered state as inducing a longitudinal field \cite{coldea,lee}.
Further test of the $E_8$ description would be provided by measuring the spin dynamics
at finite temperatures. Our study here provides a concrete prediction of
the temperature-dependence of the NMR relaxation rate in the $E_8$ model, which can be used for
 the desired further test.
During the final stage of writing the present manuscript, NMR measurements
in CoNb$_2$O$_6$ have been reported in the higher-temperature quantum critical regime \cite{kinross};
such measurements
at the lower-temperature $E_8$ regime  should therefore be feasible.

To summarize, we have determined the local dynamical spin structure factor of the
perturbed quantum-critical
Ising chain at temperatures and frequencies that are small compared to the mass of the
lightest $E_8$ particle. The frequency dependence
shows a logarithmic singularity. Our calculation yields a concrete prediction for the temperature
dependence of the NMR relaxation
rate, which we have suggested as a means to further test the $E_8$ description of the spin
dynamics in CoNb$_2$O$_6$.

\section{Acknowledgement}
We thank R. Coldea, F.H.L.Essler, G\'abor Tak\'acs and J. H.H. Perk  for useful discussions.
J.W.\ and Q.S.\ acknowledge the support provided in part by
 the NSF Grant No.\ DMR-1309531 and the Robert A.\ Welch Foundation Grant No.\ C-1411, and M.K.\ acknowledges that  by the Marie Curie IIF Grant PIIF-GA-2012- 330076. Q.S.\  also acknowledges the support of the Alexander von Humboldt Foundation, and the hospitality
of the the Karlsruhe Institute of Technology and
the Institute of Physics of Chinese Academy of Sciences.

\newpage
\onecolumngrid
\setcounter{figure}{0}
\makeatletter
\renewcommand{\thefigure}{S\@arabic\c@figure}
\setcounter{equation}{0} \makeatletter
\renewcommand \theequation{S\@arabic\c@equation}
\section*{{\Large Supplementary Material ---- Finite temperature spin dynamics \\ in a perturbed quantum critical Ising chain with an $E_8$ symmetry}}
{\hfill Jinda Wu, M\'arton Kormos and Qimiao Si \hfill}
\vskip 1.0 cm

\section{D\MakeLowercase{erivation} \MakeLowercase{of} $\chi_{yy}(x,t)$ }
The Hamiltonian of one dimensional transverse field Ising model with a longitudinal field can be expressed as
\be
H =  - J\left( {\sum\limits_i {\sigma _i^z \sigma _{i + 1}^z }  + g\sum\limits_i {\sigma _i^x }  + h_z\sum\limits_i {\sigma _i^z } } \right)
\ee
where $J$, $g$ and $h_z$ have the same meaning as in the main text. Consider $C\left( {i,j,t,T} \right) = \left\langle {\sigma _i^z (t)\sigma _j^z (0)} \right\rangle _T$, where $\left\langle  \cdots  \right\rangle _T$ denotes thermal averaging. We have
\be
\frac{{\partial C\left( {i,j,t,T} \right)}}{{\partial t}} =  - Jgi\left\langle {e^{iHt} \left[ {\sigma _i^x (0),\;\sigma _i^z (0)} \right]e^{ - iHt} \sigma _i^z (0)} \right\rangle  =  - 2Jg\left\langle {\sigma _i^y (t)\sigma _j^z (0)} \right\rangle _T  =  - 2Jg\left\langle {\sigma _i^y (0)\sigma _j^z ( - t)} \right\rangle _T
\ee
and
\be
\frac{{\partial ^2 C\left( {i,j,t,T} \right)}}{{\partial t^2 }} = \left( { - Jg} \right)2\frac{{\partial \left\langle {\sigma _i^y (0)\sigma _j^z ( - t)} \right\rangle _T }}{{\partial t}} = \left( { - Jg} \right)^2 ( - 4)\left\langle {\sigma _i^y (t)\sigma _j^y (0)} \right\rangle _T  =  - 4\left( {Jg} \right)^2 \left\langle {\sigma _i^y (t)\sigma _j^y (0)} \right\rangle _T
\ee
Recall the definition of linear response $\chi _{ij,T}^{\alpha \alpha }, \alpha=x,y,z$,
\be
\chi _{ij,T}^{\alpha \alpha }  =  - i\theta (t)\left\langle {\left[ {\sigma _i^\alpha  (t),\sigma _j^\alpha  (0)} \right]} \right\rangle _T
\ee
We have
\be
\chi _T^{yy} (x,t) =  - \frac{1}{{4(gJ)^2 }}\frac{{\partial ^2 \chi _T^{zz} (x,t)}}{{\partial t^2 }}
\ee
Then
\be
\chi _T^{yy} (\omega ) = \frac{{\omega ^2 }}{{4(gJ)^2 }}\chi _T^{zz} (\omega )
\ee

\section{R\MakeLowercase{elevant}\;F\MakeLowercase{orm} F\MakeLowercase{actors} u\MakeLowercase{sed} \MakeLowercase{in} \MakeLowercase{the} M\MakeLowercase{ain} T\MakeLowercase{ext}}
The main text considered two- and three-particle form factors of the $E_8$ model. The relevant two-particle form factors are known in the literature\cite{delfino1,delfino2}. Here, for completeness, we present their detailed expressions, where``$n$" in $F_n^\sigma$ is explicitly written as types of particles it contains.
\be
F_{aa}^\sigma  \left( {\theta _1 ,\theta _2 } \right) = \left\{ {c_{11}^0  + c_{11}^1 \cosh \left( {\theta _1  - \theta _2 } \right)} \right\}\left\{ { - i\sinh \left( {\frac{{\theta _1  - \theta _2 }}{2}} \right)} \right\}\frac{{T_{2/3} \left( {\theta _1  - \theta _2 } \right)T_{2/5} \left( {\theta _1  - \theta _2 } \right)T_{1/15} \left( {\theta _1  - \theta _2 } \right)}}{{P_{2/3} \left( {\theta _1  - \theta _2 } \right)P_{2/5} \left( {\theta _1  - \theta _2 } \right)P_{1/15} \left( {\theta _1  - \theta _2 } \right)}} \ee
\bea
&& F_{bb}^\sigma  \left( {\theta _1 ,\theta _2 } \right) = \left\{ {c_{22}^0  + c_{22}^1 \cosh \left( {\theta _1  - \theta _2 } \right) + c_{22}^2 \cosh ^2 \left( {\theta _1  - \theta _2 } \right) + c_{22}^3 \cosh ^3 \left( {\theta _1  - \theta _2 } \right)} \right\}\left\{ { - i\sinh \left( {\frac{{\theta _1  - \theta _2 }}{2}} \right)} \right\}
 \cdot \;\; \nonumber   \\
&\;&\;\;\;\;\;\;\;\;\; \cdot \frac{{T_{4/5} \left( {\theta _1  - \theta _2 } \right)T_{2/3} \left( {\theta _1  - \theta _2 } \right)T_{7/15} \left( {\theta _1  - \theta _2 } \right)T_{4/15} \left( {\theta _1  - \theta _2 } \right)T_{1/15} \left( {\theta _1  - \theta _2 } \right)\left( {T_{2/5} \left( {\theta _1  - \theta _2 } \right)} \right)^2 }}{{P_{4/5} \left( {\theta _1  - \theta _2 } \right)P_{2/3} \left( {\theta _1  - \theta _2 } \right)P_{7/15} \left( {\theta _1  - \theta _2 } \right)P_{4/15} \left( {\theta _1  - \theta _2 } \right)P_{1/15} \left( {\theta _1  - \theta _2 } \right)P_{2/5} \left( {\theta _1  - \theta _2 } \right)P_{3/5} \left( {\theta _1  - \theta _2 } \right)}}
\eea
\bea
F_{cc}^\sigma  \left( {\theta _1 ,\theta _2 } \right) = \left\{ {c_{33}^0  + c_{33}^1 \cosh \left( {\theta _1  - \theta _2 } \right) + c_{33}^2 \cosh ^2 \left( {\theta _1  - \theta _2 } \right) + c_{33}^3 \cosh ^3 \left( {\theta _1  - \theta _2 } \right) + c_{33}^4 \cosh ^4 \left( {\theta _1  - \theta _2 } \right)} \right\} \cdot  \nonumber \\
  \cdot \left\{ { - i\sinh \left( {\frac{{\theta _1  - \theta _2 }}{2}} \right)} \right\}T_{11/30} \left( {\theta _1  - \theta _2 } \right)\left[ {T_{2/3} \left( {\theta _1  - \theta _2 } \right)} \right]^3 T_{7/15} \left( {\theta _1  - \theta _2 } \right)\left[ {T_{2/5} \left( {\theta _1  - \theta _2 } \right)} \right]^3 T_{2/15} \left( {\theta _1  - \theta _2 } \right) \cdot \nonumber \\
  \cdot \left[ {T_{1/15} \left( {\theta _1  - \theta _2 } \right)} \right]^2 /\left\{ {P_{11/30} \left( {\theta _1  - \theta _2 } \right)P_{7/15} \left( {\theta _1  - \theta _2 } \right)P_{2/15} \left( {\theta _1  - \theta _2 } \right)\left[ {P_{2/3} \left( {\theta _1  - \theta _2 } \right)} \right]^2 P_{1/3} \left( {\theta _1  - \theta _2 } \right) \cdot } \right. \nonumber\\
 \left. { \cdot \left[ {P_{2/5} \left( {\theta _1  - \theta _2 } \right)} \right]^2 P_{3/5} \left( {\theta _1  - \theta _2 } \right)P_{1/15} \left( {\theta _1  - \theta _2 } \right)P_{14/15} \left( {\theta _1  - \theta _2 } \right)} \right\}
 \eea
 \bea
F_{ab}^\sigma  \left( {\theta _1 ,\theta _2 } \right) &=& \left\{ {c_{12}^0  + c_{12}^1 \cosh \left( {\theta _1  - \theta _2 } \right) + c_{12}^2 \cosh ^2 \left( {\theta _1  - \theta _2 } \right)} \right\}\frac{{T_{4/5} \left( {\theta _1  - \theta _2 } \right)T_{3/5} \left( {\theta _1  - \theta _2 } \right)T_{7/15} \left( {\theta _1  - \theta _2 } \right)T_{4/15} \left( {\theta _1  - \theta _2 } \right)}}{{P_{4/5} \left( {\theta _1  - \theta _2 } \right)P_{3/5} \left( {\theta _1  - \theta _2 } \right)P_{7/15} \left( {\theta _1  - \theta _2 } \right)P_{4/15} \left( {\theta _1  - \theta _2 } \right)}}\nonumber \\ \\
 F_{ac}^\sigma  \left( {\theta _1 ,\theta _2 } \right) &=& \left\{ {c_{13}^0  + c_{13}^1 \cosh \left( {\theta _1  - \theta _2 } \right) + c_{13}^2 \cosh ^2 \left( {\theta _1  - \theta _2 } \right) + c_{13}^3 \cosh ^3 \left( {\theta _1  - \theta _2 } \right)} \right\} \cdot  \nonumber \\
 &&  \cdot \frac{{T_{29/30} \left( {\theta _1  - \theta _2 } \right)T_{7/10} \left( {\theta _1  - \theta _2 } \right)T_{13/30} \left( {\theta _1  - \theta _2 } \right)T_{1/10} \left( {\theta _1  - \theta _2 } \right)\left[ {T_{11/30} \left( {\theta _1  - \theta _2 } \right)} \right]^2 }}{{P_{29/30} \left( {\theta _1  - \theta _2 } \right)P_{7/10} \left( {\theta _1  - \theta _2 } \right)P_{13/30} \left( {\theta _1  - \theta _2 } \right)P_{1/10} \left( {\theta _1  - \theta _2 } \right)P_{11/30} \left( {\theta _1  - \theta _2 } \right)P_{19/30} \left( {\theta _1  - \theta _2 } \right)}} \\
 F_{bc}^\sigma  \left( {\theta _1 ,\theta _2 } \right) &=& \left\{ {c_{23}^0  + c_{23}^1 \cosh \left( {\theta _1  - \theta _2 } \right) + c_{23}^2 \cosh ^2 \left( {\theta _1  - \theta _2 } \right) + c_{23}^3 \cosh ^3 \left( {\theta _1  - \theta _2 } \right) + c_{23}^4 \cosh ^4 \left( {\theta _1  - \theta _2 } \right)} \right\} \cdot  \nonumber \eea
\vspace{-7mm}
 \be
\frac{{T_{25/30} \left( {\theta _1  - \theta _2 } \right)T_{19/30} \left( {\theta _1  - \theta _2 } \right)T_{9/30} \left( {\theta _1  - \theta _2 } \right)\left[ {T_{7/30} \left( {\theta _1  - \theta _2 } \right)} \right]^2 \left[ {T_{13/30} \left( {\theta _1  - \theta _2 } \right)} \right]^2 T_{15/30} \left( {\theta _1  - \theta _2 } \right)}}{{P_{25/30} \left( {\theta _1  - \theta _2 } \right)P_{19/30} \left( {\theta _1  - \theta _2 } \right)P_{9/30} \left( {\theta _1  - \theta _2 } \right)P_{7/30} \left( {\theta _1  - \theta _2 } \right)P_{23/30} \left( {\theta _1  - \theta _2 } \right)P_{13/30} \left( {\theta _1  - \theta _2 } \right)P_{17/30} \left( {\theta _1  - \theta _2 } \right)P_{15/30} \left( {\theta _1  - \theta _2 } \right)}}
\ee
where
\be
T_\lambda  (\theta ) = \exp \left\{ {2\int_0^\infty  {\frac{{dt}}{t}\frac{{\cosh \left[ {\left( {\lambda  - 1/2} \right)t} \right]}}{{\cosh \left( {t/2} \right)\sinh t}}\sin ^2 \frac{{\left( {i\pi  - \theta } \right)t}}{{2\pi }}} } \right\}\;\;and\;\;P_\lambda  (\theta ) = \frac{{\cos \left( {\lambda \pi } \right) - \cosh \theta }}{{2\cos ^2 \left( {\lambda \pi /2} \right)}} .
\ee
The coefficients $\{c_{ij}^k\}$ in the above expressions are
\cite{delfino2}:
\[
\begin{array}{l}
 c_{11}^0 ,c_{11}^1  =  - 10.19307727, - 2.09310293 \\
 c_{22}^0 ,c_{22}^1 ,c_{22}^2 ,c_{22}^3  =  - 500.2535896, - 791.3745549, - 338.8125724, - 21.48559881 \\
 c_{33}^0 ,c_{33}^1 ,c_{33}^2 ,c_{33}^3 ,c_{33}^4 ,c_{33}^5  =  - 87821.70785, - 267341.1276, - 301093.9432, - 150512.4122, - 30166.99117, - 1197.056497 \\
 c_{12}^0 ,c_{12}^1 ,c_{12}^2  =  - {\rm{70}}{\rm{.29218939519}}, - {\rm{71}}{\rm{.792063506}}, - {\rm{7}}{\rm{.9790221816}} \\
 c_{13}^0 ,c_{13}^1 ,c_{13}^2 ,c_{13}^3  =  - {\rm{7049}}{\rm{.622303}}, - {\rm{13406}}{\rm{.48877}}, - {\rm{6944}}{\rm{.416956}}, - {\rm{582}}{\rm{.2557366}} \\
 c_{23}^0 ,c_{23}^1 ,c_{23}^2 ,c_{23}^3 ,c_{23}^4  =  - {\rm{3579}}{\rm{.556465}}, - {\rm{8436}}{\rm{.850081}}, - {\rm{6618}}{\rm{.297073}}, - {\rm{1846}}{\rm{.579035}}, - {\rm{92}}{\rm{.73452314}} \\
 \end{array}
\]

The relevant three-particle form factor of the $E_8$ model is,
\be
F_{aaa}^\sigma  \left( {\theta _1 ,\theta _2 ,\theta _3 } \right) = Q_{aaa}^\sigma  \left( {\theta _1 ,\theta _2 ,\theta _3 } \right)\frac{{F_{aa}^{\min } \left( {\theta _1  - \theta _2 } \right)}}{{\left( {e^{\theta _1 }  + e^{\theta _2 } } \right)D_{aa} \left( {\theta _1  - \theta _2 } \right)}}\frac{{F_{aa}^{\min } \left( {\theta _1  - \theta _3 } \right)}}{{\left( {e^{\theta _1 }  + e^{\theta _3 } } \right)D_{aa} \left( {\theta _1  - \theta _3 } \right)}}\frac{{F_{aa}^{\min } \left( {\theta _2  - \theta _3 } \right)}}{{\left( {e^{\theta _2 }  + e^{\theta _3 } } \right)D_{aa} \left( {\theta _2  - \theta _3 } \right)}}
\ee
where\cite{delfino1,delfino2}
\be
\frac{{F_{aa}^{\min } \left( {\theta _i  - \theta _j } \right)}}{{D_{aa} \left( {\theta _i  - \theta _j } \right)}} = \left\{ { - i\sinh \left( {\frac{{\theta _i  - \theta _j }}{2}} \right)} \right\}\frac{{T_{2/3} \left( {\theta _i  - \theta _j } \right)T_{2/5} \left( {\theta _i  - \theta _j } \right)T_{1/15} \left( {\theta _i  - \theta _j } \right)}}{{P_{2/3} \left( {\theta _i  - \theta _j } \right)P_{2/5} \left( {\theta _i  - \theta _j } \right)P_{1/15} \left( {\theta _i  - \theta _j } \right)}}
\ee
and\cite{delfino3}
\bea
&& Q_{aaa}^\sigma  \left( {\theta _1 ,\theta _2 ,\theta _3 } \right)=1148.690509{\rm{ }}e^{3\theta _1 }  + 46.76252978{\rm{ }}e^{4\theta _1  - \theta _2 } + 1148.690509{\rm{ }}e^{3\theta _2 }  + 3703.911733{\rm{ }}e^{2\theta _1  + \theta _2 } \nonumber \\
&&  + 3703.911733{\rm{ }}e^{\theta _1  + 2\theta _2 }  + 46.76252978{\rm{ }}e^{ - \theta _1  + 4\theta _2 } + 4.354182251{\rm{ }}e^{4\theta _1  + \theta _2  - 2\theta _3 }  + 46.76252978{\rm{ }}e^{3\theta _1  + 2\theta _2  - 2\theta _3 } \nonumber \\
&&  + 46.76252978{\rm{ }}e^{2\theta _1  + 3\theta _2  - 2\theta _3 }  + 4.354182251{\rm{ }}e^{\theta _1  + 4\theta _2  - 2\theta _3 } + 46.76252978{\rm{ }}e^{4\theta _1  - \theta _3 }  + 604.2577928{\rm{ }}e^{3\theta _1  + \theta _2  - \theta _3 } \nonumber \\
&&  + 1148.690509{\rm{ }}e^{2\theta _1  + 2\theta _2  - \theta _3 }  + 604.2577928\;e^{\theta _1  + 3\theta _2  - \theta _3 } + 46.76252978{\rm{ }}e^{4\theta _2  - \theta _3 }  + 1148.690509{\rm{ }}e^{3\theta _3 }\nonumber  \\
&&  + 3703.911733{\rm{ }}e^{2\theta _1  + \theta _3 }  + 4.354182251{\rm{ }}e^{4\theta _1  - 2\theta _2  + \theta _3 } + 604.2577928{\rm{ }}e^{3\theta _1  - \theta _2  + \theta _3 }  + 6286.815608\;e^{\theta _1  + \theta _2  + \theta _3 } \nonumber \\
&&  + 3703.911733{\rm{ }}e^{2\theta _2  + \theta _3 }  + 604.2577928\;e^{ - \theta _1  + 3\theta _2  + \theta _3 } + 4.354182251{\rm{ }}e^{ - 2\theta _1  + 4\theta _2  + \theta _3 }  + 3703.911733{\rm{ }}e^{\theta _1  + 2\theta _3 }  \nonumber\\
&&  + 46.76252978{\rm{ }}e^{3\theta _1  - 2\theta _2  + 2\theta _3 }  + 1148.690509{\rm{ }}e^{2\theta _1  - \theta _2  + 2\theta _3 }  + 3703.911733{\rm{ }}e^{\theta _2  + 2\theta _3 }  + 1148.690509{\rm{ }}e^{ - \theta _1  + 2\theta _2  + 2\theta _3 }  \nonumber\\
&&  + 46.76252978{\rm{ }}e^{ - 2\theta _1  + 3\theta _2  + 2\theta _3 }  + 46.76252978{\rm{ }}e^{2\theta _1  - 2\theta _2  + 3\theta _3 } + 604.2577928{\rm{ }}e^{\theta _1  - \theta _2  + 3\theta _3 }  + 604.2577928{\rm{ }}e^{ - \theta _1  + \theta _2  + 3\theta _3 } \nonumber \\
&&  + 46.76252978{\rm{ }}e^{ - 2\theta _1  + 2\theta _2  + 3\theta _3 }  + 46.76252978{\rm{ }}e^{ - \theta _1  + 4\theta _3 }  + 4.354182251{\rm{ }}e^{\theta _1  - 2\theta _2  + 4\theta _3 }  + 46.76252978{\rm{ }}e^{ - \theta _2  + 4\theta _3 }  \nonumber\\
&&  + 4.354182251{\rm{ }}e^{ - 2\theta _1  + \theta _2  + 4\theta _3 }
\eea

\section{D\MakeLowercase{erivation}\;\MakeLowercase{of}\;$S_{1,1} (\omega ,q)$\;(E\MakeLowercase{q}.(19) \MakeLowercase{of} \MakeLowercase{the} M\MakeLowercase{ain} T\MakeLowercase{ext}) }

We calculate $S_{1,1}$ using the finite volume
regularization scheme \cite{pozsgay,szecsenyi}.
We have $D_{00} = \left\langle \sigma  \right\rangle _0^2$, and
\be
\begin{gathered}
 D_{11} (x,t ) = \int_{ - \infty }^\infty  {\frac{{d\theta _1 }}{{2\pi }}\int_{ - \infty }^\infty  {\frac{{d\theta' _1 }}{{2\pi }}F_2^\sigma  \left( {\theta_1  + i\pi ,\theta' _1} \right)F_2^\sigma  \left( {\theta' _1 + i\pi ,\theta_1 } \right)} }\cdot  \\
\cdot e^{ - \beta \Delta _1 \cosh \theta _1 } e^{ - i\left( {\Delta _1 \sinh \theta _1  - \Delta _2 \sinh \theta ' _1 } \right)x} e^{ - i \left( {\Delta _2 \cosh \theta ' _1  - \Delta _1 \cosh \theta _1 } \right) t } + 2\left\langle \sigma  \right\rangle _0 F_{2s}^\sigma  \int_{ - \infty }^\infty  {\frac{{d\theta }}{{2\pi }}e^{ - \beta \Delta _1 \cosh \theta _1 } } \\
\end{gathered}
\ee
From now until the calculation of $D_{22}$, we will focus on the time-dependent parts, i.e., the connected pieces of the correlation functions.
The time-independent parts, i.e., the disconnected pieces, will be discussed after the analysis on the time-dependent parts of $D_{22}$. We then have
\be
\begin{gathered}
S_{1,1} (q,\omega ) =    \int {d\theta _1 d\theta ' _1 F_2^\sigma  (\theta _1  + i\pi ,\theta ' _1 )F_2^\sigma  (\theta ' _1  + i\pi ,\theta _1 ) e^{ - \beta \Delta _1 \cosh \theta _1 } \cdot } \;\;\;\;\;\;\;\;\;\;\;\;\;\;\;\;\;\;\;\;\;\;\;\;\;\;\;\;\;\;\;\;\;\;\;\; \\
\cdot \delta (q + \Delta _1 \sinh \theta _1  - \Delta _2 \sinh \theta ' _1 ) \delta (\omega  + \Delta _1 \cosh \theta _1  - \Delta _2 \cosh \theta ' _1 ) \\
\end{gathered}
\ee
Denote
\be
\left\{ \begin{array}{l}
 y = \Delta _1 \sinh \theta _1  - \Delta _2 \sinh \theta ' _1  \\
 z = \Delta _1 \cosh \theta _1  - \Delta _2 \cosh \theta ' _1  \\
 \end{array} \right.
\ee
then
\be
d\theta _1 d\theta ' _1  = \left| {\frac{{\partial \left( {\theta _1 ,\theta ' _1 } \right)}}{{\partial \left( {y,z} \right)}}} \right|dydz = \frac{{dydz}}{{\Delta _1 \Delta _2 |\sinh \alpha| }},
\ee
where
\be
 \alpha  = {\rm{arccosh}} \left[ \frac{{\Delta _1^2  + \Delta _2^2  - \left( {\omega ^2  - q^2 } \right)}}{{2\Delta _1 \Delta _2 }} \right] .
\ee
Noticing that the integration ranges for new variables $y$ and $z$ run from $-\infty$ to $+\infty$, we can easily perform the integral in the structure factor and find
\bea
&& S_{1,1} (q,\omega ) =   \int {d\theta _1 d\theta'_1 F_2^\sigma  (\theta _1  + i\pi ,\theta ' _1 )F_2^\sigma  (\theta' _1  + i\pi ,\theta _1 )\delta (q + \Delta _1 \sinh \theta _1  - \Delta _2 \sinh \theta ' _1 ) } \cdot \nonumber  \\
 &&\;\;\;\;\;\;\;\;\;\;\;\;\;\;\;\;\;\;\;\;\;\;\;\;\;\;\;\;\;\;\;\;\;\;\;\;\;\;\;\;\;\;\;\;\;\;\;\;\;\;\;\;\;\;\;\;\;\;\;\;\;\;\;\;\;\;\;\cdot\delta (\omega  + \Delta _1 \cosh \theta _1  - \Delta _2 \cosh \theta ' _1 )e^{ - \beta \Delta _1 \cosh \theta _1 } \label{s11qomega} \\
  &=&   \left. {\frac{{F_2^\sigma  (\theta _1  + i\pi ,\theta ' _1 )F_2^\sigma  (\theta' _1  + i\pi ,\theta _1 )e^{ - \beta \Delta _2 \cosh \theta ' _1 } e^{\beta \omega } }}{{\Delta _1 \Delta _2 |\sinh \alpha| }}} \right|_{\theta _1  - \theta ' _1  = \alpha }  + \left. {\frac{{F_2^\sigma  (\theta _1  + i\pi ,\theta ' _1 )F_2^\sigma  (\theta ' _1  + i\pi ,\theta _1 )e^{ - \beta \Delta _2 \cosh \theta ' _1 } e^{\beta \omega } }}{{\Delta _1 \Delta _2 |\sinh \alpha| }}} \right|_{\theta _1  - \theta ' _1  =  - \alpha } \label{dynamics1100}\\
  &=& \frac{{\left| {F_2^\sigma  (\alpha + i\pi,0 )} \right|^2 e^{ - \beta \Delta _1 \cosh \theta _{1_+} }  }}{{\Delta _1 \Delta _2 |\sinh \alpha| }}  +  \frac{{\left| {F_2^\sigma  (-\alpha  + i\pi,0 )} \right|^2 e^{ - \beta \Delta _1 \cosh \theta _{1_-} }  }}{{\Delta _1 \Delta _2 |\sinh \alpha|}} \label{dynamics110} \\
  &=& \frac{{\left| {F_2^\sigma  (\alpha + i\pi, 0 )} \right|^2 \left(e^{ - \beta \Delta _1 \cosh \theta _{1_+} } +e^{ - \beta \Delta _1 \cosh \theta _{1_-} } \right) }}{{\Delta _1 \Delta _2 |\sinh \alpha| }}
     \label{dynamics11}
\eea
where
\bea
 \cosh \theta _{1_ \pm  }  = \frac{{\omega \Delta _1  - \omega \Delta _2 \Gamma  \pm q\Delta _2 \sqrt {\Gamma ^2  - 1} }}{{q^2  - \omega ^2 }}
 \eea
In going from Eq.~(\ref{dynamics1100}) to
 Eq.~(\ref{dynamics110}), we have used
the fact that the form factor is only dependent on the difference between any two rapidities. We then recover Eq.(19) of the main text.

\section{C\MakeLowercase{alculation}\;\MakeLowercase{of}\;$S_{1,1} (\omega ,q)$\;(E\MakeLowercase{q}.(21)
\MakeLowercase{of} \MakeLowercase{the} M\MakeLowercase{ain} T\MakeLowercase{ext}) \MakeLowercase{for} E\MakeLowercase{qual} M\MakeLowercase{asses} \MakeLowercase{at} L\MakeLowercase{ow} F\MakeLowercase{requencies} }
In Eq.~(\ref{s11qomega}) above, the $q$ integration followed by  the $\theta'_1$ integration gives rise to
\be
S_{1,1} (\omega ) = \frac{2}
{{\Delta _i }}\int {d\theta _1 \frac{{\left( {|F_2^\sigma  (\theta _1  - \theta '_{1 + } (\theta _1 ,\omega ) + i\pi ,0)|^2  + |F_2^\sigma  (\theta _1  - \theta '_{1 - } (\theta _1 ,\omega ) + i\pi ,0)|^2 } \right)e^{ - \beta \Delta _i \cosh \theta _1 } }}
{{\sqrt {\left( {\cosh \theta _1  + \omega /\Delta _i } \right)^2  - 1} }}}
\ee
where $\theta '_{1 \pm }  =  \pm {\text{arccosh}}\left( {\cosh \theta _1  + \omega /\Delta _i } \right)$. We make a further variable transform $\cosh \theta _1  = x - \omega _i$ ($\omega_i = \omega/(2\Delta_i)$)  and have
\bea
 \left. {S_{1,1} (\omega_i )} \right|_{\Delta _1  = \Delta _2  = \Delta _i }  = \frac{{2e^{\omega /(2T)} }}{\Delta_i }\int_{1 + \omega_i }^\infty  {dx\frac{{(F\left( {\omega_i ,q_-(x)} \right)+F\left( {\omega_i ,q_+(x)} \right))\exp \left\{ { - \frac{{\Delta _i }}{T}x} \right\}}}{{\sqrt {\left( {x + 1 + \omega_i } \right)\left( {x + 1 - \omega_i } \right)\left( {x - 1 + \omega_i } \right)\left( {x - 1 - \omega_i } \right)} }}} \label{S11equalmass}
  \eea
where $\omega _i  = \omega /(2\Delta _i )$, and
\bea
F(\omega_i ,q_{\pm}(x)) =\left| {F_{ii}^\sigma  ({\rm{arccosh}}\left[ {{\rm{1 + }}\left( { - \omega_i ^2  + x^2  - 1 \pm \sqrt {\left( {x - 1 - \omega_i } \right)\left( {x - 1 + \omega_i } \right)\left( {x + 1 + \omega_i } \right)\left( {x + 1 - \omega_i } \right)} } \right)} \right] + i\pi, 0 )} \right|^2
  \eea
We can also get Eq.~(\ref{S11equalmass}) by making variable transform $x =  - \omega /(2\Delta _i ) + \left[ {q/\left( {2\Delta _i } \right)} \right]\sqrt {\left( {q^2  - \omega ^2  + 4\Delta _i^2 } \right)/\left( {q^2  - \omega ^2 } \right)}$ for the $q$ integration over  Eq.~(\ref{dynamics11}). The exponential-decaying factor in the integrand of Eq.~(\ref{S11equalmass}) indicates that the dominant contribution come from the regime where $x$ is close to $1+\omega_i$. Since $\omega_i$ is small, in this regime we can approximate $F(\omega_i ,q_{\pm}(x))$ as
\be
F(\omega_i ,q_{\pm}(x)) \approx |F_{ii}^\sigma(i\pi,0)|^2 \;\; (i=a,b,c,d,e,f,g,h).
\ee
Then we have
\bea
&& \left. {S_{1,1} (\omega \to 0,\omega/\Delta_i \ll 1 )} \right|_{\Delta _1  = \Delta _2  = \Delta _i } \\
 &\approx& \frac{{4e^{\omega /(2T)} }}{\Delta_i }\int_{1 + \omega_i }^\infty  {dx\frac{{\left| {F_{ii} \left( {i\pi,0 } \right)} \right|^2 \exp \left\{ { - \frac{{\Delta _i }}{T}x} \right\}}}{{\sqrt {\left( {x + 1 + \omega_i } \right)\left( {x + 1 - \omega_i } \right)\left( {x - 1 + \omega_i } \right)\left( {x - 1 - \omega_i } \right)} }}}   \\
 &\approx&  \left\{ \begin{array}{l}
  \frac{{2e^{\omega /(2T)} \left| {F_{ii} \left( {i\pi,0 } \right)} \right|^2 }}{\Delta_i }e^{ - \Delta_i /T} \left\{ {-\ln \frac{\omega }{{4T}} -\gamma _E  +\cdots \cdots} \right\} \;(\omega \ll T  \ll \Delta_i ) \\
 \frac{{2e^{ - \Delta _i /T} \left| {F_{ii} \left( {i\pi } \right)} \right|^2 }}{{\Delta _i }}\left\{ {\sqrt {\frac{{\pi T}}{\omega }}  - \frac{{\sqrt \pi  }}{4}\left( {\frac{T}{\omega }} \right)^{3/2}  +  \cdots  \cdots } \right\}\;(T \ll \omega  \ll \Delta_i ) \\
 \end{array} \right.
 \label{s11leading}
\eea
where $\left| {F_{aa}^\sigma \left( {i\pi,0 } \right)} \right|^2  \approx 65$.

\section{D\MakeLowercase{erivation}\;\MakeLowercase{of}\;$S_{1,2} (\omega ,q)$\;(E\MakeLowercase{q}.(22)
\MakeLowercase{of} \MakeLowercase{the} M\MakeLowercase{ain} T\MakeLowercase{ext}) }
We again use the finite volume
regularization scheme \cite{pozsgay,szecsenyi}, and have
\bea
&& D_{12} (x,t ) = C_{12}  - Z_1 C_{01}  \\
  &=& \frac{1}{2}\int_{C_ +  } {\frac{{d\theta _1 }}{{2\pi }}\int_{ - \infty }^\infty  {\frac{{d\theta '_1 }}{{2\pi }}\int_{ - \infty }^\infty  {\frac{{d\theta '_2 }}{{2\pi }}F_3^\sigma  \left( {\theta _1  + i\pi ,\theta '_1 ,\theta '_2 } \right)F_3^\sigma  \left( {\theta _1  + i\pi ,\theta '_2 ,\theta '_1 } \right)} } }  \\
&& \;\;\;\;\;\;\;\;\;\;\;\;\;\;\;\;\;\;\;\;\;\;\;\;\;\;\;\;\;\;\;\;\;\;\;\;\;\;\;\;\;\;e^{ - \beta \Delta _a \cosh \theta _1 } e^{ - ix\Delta _a \left( {\sinh \theta _1  - \sinh \theta '_1  - \sinh \theta '_2 } \right)} e^{ - i t \Delta _a \left( {\cosh \theta '_1  + \cosh \theta '_2  - \cosh \theta _1 } \right)}  \\
  &&+ \int_{ - \infty }^\infty  {\frac{{d\theta '_1 }}{{2\pi }}\int_{ - \infty }^\infty  {\frac{{d\theta '_2 }}{{2\pi }}\left\{ {e^{ - \beta \Delta _a \cosh \theta '_1 } e^{i\Delta _a x\sinh \theta '_2 } e^{ - it \Delta _a  \cosh \theta '_2 } S\left( {\theta '_2  - \theta '_1 } \right)F_1^\sigma  F_{3c}^\sigma  \left( {\theta '_1 |\theta '_1 \theta '_2 } \right) + \left( {\theta '_1  \leftrightarrow \theta '_2 } \right)} \right\}} }  \\
  &&- \left( {F_1^\sigma  } \right)^2 \int_{ - \infty }^\infty  {\frac{{d\theta '_1 }}{{2\pi }}\int_{ - \infty }^\infty  {\frac{{d\theta '_2 }}{{2\pi }}e^{ - \beta \Delta _a \cosh \theta '_1 } e^{i\Delta _a x\sinh \theta '_2 } e^{ - it \Delta _a \cosh \theta '_2 }  \cdot } }  \\
&& \;\;\;\;\;\;\;\;\;\;\;\;\;\;\;\;\;\;\;\;\;\;\;\;\;\;\;\;\;\;\;\;\;\;\;\;\;\;\;\;\left( {\Delta _a x\cosh \theta '_1  + \Delta _a \left( {i\beta  + t } \right)\sinh \theta '_1 } \right)\left[ {S\left( {\theta '_1  - \theta '_2 } \right) - 1} \right] \\
 && - \left( {F_1^\sigma  } \right)^2 \int_{ - \infty }^\infty  {\frac{{d\theta '_2 }}{{2\pi }}e^{ - \beta \Delta _a \cosh \theta '_2 } e^{i\Delta _a x\sinh \theta '_2 } e^{ - it \Delta _a \cosh \theta '_2 } }
  \eea
where $C_+$ is used to denote the integration contour from $-\infty$ to $\infty$ slightly above the real axis on the rapidity complex plane, and\cite{pozsgay,szecsenyi}
\be
F_3^\sigma  (\theta _1  + i\pi ,\theta '_1 ,\theta '_2 ) = \frac{{i\left( {1 - S(\theta '_1  - \theta '_2 )} \right)F_1^\sigma  }}
{{\theta _1  - \theta '_1 }} + \frac{{i\left( {S(\theta '_1  - \theta '_2 ) - 1} \right)F_1^\sigma  }}
{{\theta _1  - \theta '_2 }} + F_{3rc}^\sigma  \left( {\theta _1  + i\pi |\theta '_1 ,\theta _2 } \right) \label{f3}
\ee
where $S_{aa}$ is the scattering matrix for $a-a$ channel, and  $F_{3rc}^\sigma  \left( {\theta _1  + i\pi |\theta '_1 ,\theta _2 } \right)$ is regular on real axis.

For $x=0$, it's easy to see that the last three terms do not contribute to low-frequency ($\omega \ll \Delta_a$) response of local DSF. Fom the first integration we have
\bea
 S_{1,2} (q,\omega ) &=& \frac{1}{2}\int {\frac{{d\theta _1 }}{{2\pi }}\int_{ - \infty }^\infty  {d\theta '_1 \int_{ - \infty }^\infty  {d\theta '_2 F_3^\sigma  \left( {\theta _1  + i\pi +i\varepsilon ,\theta '_1 ,\theta '_2 } \right)F_3^\sigma  \left( {\theta _1  + i\pi + i\varepsilon ,\theta '_2 ,\theta '_1 } \right)} } }  \\
 &&\delta \left( {q + \Delta _a \sinh \theta _1  - \Delta _a \left( {\sinh \theta '_1  + \sinh \theta '_2 } \right)} \right)\delta \left( {\omega  + \Delta _a \cosh \theta _1  - \Delta _a \left( {\cosh \theta '_1  + \cosh \theta '_2 } \right)} \right)
 \eea
The energy-momentum conservation yields
\be
\left\{ \begin{array}{l}
 0 = q + \sinh \theta _1  - \sinh \theta '_1  - \sinh \theta '_2  \\
 0 = \omega  + \cosh \theta _1  - \cosh \theta '_1  - \cosh \theta '_2  \\
 \end{array} \right.
\ee
For $F_{3rc}^\sigma$ we can integrate over $\theta'_1$ and $\theta'_2$, yielding (because the masses of three particles are equal to each other, $S_{21}(q, \omega)=S_{12}(q, \omega)$)
\be
S_{(1,2)+(2,1)} (q,\omega ) = \frac{1}{{2\pi }}\int_{ - \infty }^\infty  {\frac{{d\theta _1 }}{{\left| {\sqrt {\left( {f(\tilde\omega ,\tilde q,\theta _1 ) - 1} \right)^2  - 1} } \right|}}F_{3rc}^\sigma  \left( {\theta _1  + i\pi|\ln z_+,\ln z_-} \right)F_{3rc}^\sigma  \left( {\theta _1  + i\pi|\ln z_-,\ln z_+} \right)e^{ - \beta \Delta_a \cosh \theta _1 } } \label{dynamics12}
\ee
with
\be
z_\pm =
\frac{1}{2}(\tilde\omega  + \cosh \theta   + \tilde q + \sinh \theta)\left(1   \pm  2\sqrt{1-2/f(\tilde\omega,\tilde q,\theta)}\right),
\ee
and
\be
f(\tilde\omega ,\tilde q,\theta ) = \left[ {\left( {\tilde\omega  + \cosh \theta  } \right)^2  - \left( {\tilde q + \sinh \theta  } \right)^2 } \right]/2
\ee
where $\tilde \omega=\omega/\Delta_a$ and $\tilde q=q/\Delta_a$. Thus  we recover Eq.~(22) of the main text.
The energy-momentum conservation gives a constraint: $f(\tilde\omega ,\tilde q,\theta  )\geqslant 2 $, i.e., $(\tilde\omega  - \tilde q)e^{2\theta }  + (\tilde\omega ^2  - \tilde q^2  - 3)e^\theta   + \tilde\omega  + \tilde q \ge 0$. This constraint allows zero in the denominator of the integrand in Eq.~(\ref{dynamics12}), which is a branch point. This can be clearly shown after a variable transform $e^\theta \to x$ and expanding $x$ around zero. Thus the integration will smooth out the superficial singularity leaving us with a regular integration over $\theta$. Furthermore, if $\tilde\omega > \tilde q \geq 0 $, we can get the constraint for rapidity $0 < 2e^\theta   <  - \frac{{\tilde\omega ^2  - \tilde q^2  - 3}}{{\tilde\omega  - \tilde q}} - \sqrt {\left( {\frac{{\tilde\omega ^2  - \tilde q^2  - 3}}{{\tilde\omega  - \tilde q}}} \right)^2 - 4\frac{{\tilde\omega  + \tilde q}}{{\tilde\omega  - \tilde q}}}$ or $2e^\theta   >  - \frac{{\tilde\omega ^2  - \tilde q^2  - 3}}{{\tilde\omega  - \tilde q}} + \sqrt {\left( {\frac{{\tilde\omega ^2  - \tilde q^2  - 3}}{{\tilde\omega  - \tilde q}}} \right)^2 - 4\frac{{\tilde\omega  + \tilde q}}{{\tilde\omega  - \tilde q}}}$. However, it's easy to see in these two ranges that, because $\tilde\omega  \ll 1$, we will have $\cosh \theta \sim 1/\tilde\omega$, making it negligible in the zero frequency limit. If $\tilde\omega \leq \tilde q $, we get constraint on the rapidity of $\theta$ as (without loss of generality we choose $\tilde\omega > 0$): $0  <  2 e^{\theta} <  \left[\left( {\tilde\omega ^2  - \tilde q^2  - 3} \right)+\sqrt {\left( {\tilde\omega ^2  - \tilde q^2  - 1} \right)\left( {\tilde\omega ^2  - \tilde q^2  - 9} \right)}\right]/(\tilde q-\tilde\omega)  \equiv \mu(\tilde\omega,\tilde q) $. We  can then determine the maximum of $\mu(\tilde\omega,\tilde q)$ to be located at $\tilde q_m = \sqrt{(3-\tilde\omega)(1-\tilde\omega)}$. Again recalling $\tilde\omega \ll 1$, we have
$\cosh\theta \gtrsim 2-\tilde\omega$. This indicates that a small region of $\tilde q$ exists, in which $\cosh\theta$ is
slightly  smaller than $2$. Therefore, we will include in our numerical calculation the channels $D_{1,2}+D_{2,1}$.

For the leftover two parts in Eq.~(\ref{f3}), we have
\bea
  \frac{{i\left( {1 - S(\theta '_1  - \theta '_2 )} \right)F_1^\sigma  }}
{{\theta _1  - \theta '_1  + i\varepsilon }} = P\frac{{i\left( {1 - S(\theta '_1  - \theta '_2 )} \right)F_1^\sigma  }}
{{\theta _1  - \theta '_1 }} - i\pi \delta (\theta _1  - \theta '_1 ) ,\hfill \\
  \frac{{i\left( {S(\theta '_1  - \theta '_2 ) - 1} \right)F_1^\sigma  }}
{{\theta _1  - \theta '_2  + i\varepsilon }} = P\frac{{i\left( {S(\theta '_1  - \theta '_2 ) - 1} \right)F_1^\sigma  }}
{{\theta _1  - \theta '_2 }} - i\pi \delta (\theta _1  - \theta '_2 ) . \hfill
\eea
Here $P$ denotes principal value integration. The parts of $F_3^\sigma  \left( {\theta _1  + i\pi +i\varepsilon ,\theta '_1 ,\theta '_2 } \right)F_3^\sigma  \left( {\theta _1  + i\pi + i\varepsilon ,\theta '_2 ,\theta '_1 } \right)$  do not contribute: after integrating over $\theta'_1$ or $\theta'_2$, which
leaves us with $e^{-it\Delta_a\cosh\theta_1}$; since $\omega \ll \Delta_a$, it vanishes for the local low-frequency dynamics. For the parts of $F_3^\sigma  \left( {\theta _1  + i\pi +i\varepsilon ,\theta '_1 ,\theta '_2 } \right)F_3^\sigma  \left( {\theta _1  + i\pi + i\varepsilon ,\theta '_2 ,\theta '_1 } \right)$ containing $1/(\theta_1 - \theta'_{1/2} + i\varepsilon)^2$, we can finish an integration by part, which leaves us with only a simple principal-value integration. We can repeat the discussions for the part having delta function, and show that  it does not have any contribution. Consider now all the leftover parts in $F_3^\sigma  \left( {\theta _1  + i\pi +i\varepsilon ,\theta '_1 ,\theta '_2 } \right)F_3^\sigma  \left( {\theta _1  + i\pi + i\varepsilon ,\theta '_2 ,\theta '_1 } \right)$. Since they are all principal-value-type integration, they do not encounter any singularity. Following the discussion on the integration containing integrand of $F_{3rc}^\sigma  \left( {\theta _1  + i\pi|\ln x_+,\ln x_-} \right)F_{3rc}^\sigma  \left( {\theta _1  + i\pi|\ln x_-,\ln x_+} \right)$, they will have similar contributions
as those for Eq.~(\ref{dynamics12}). They
will likewise be included in our numerical calculations.

\section{C\MakeLowercase{alculation}\;\MakeLowercase{of}\;$D_{22}$}
Using the finite volume
regularization scheme \cite{pozsgay,szecsenyi}, we have
\be
D_{aa,aa} \left( {x,t} \right) = C_{22}  - Z_1 C_{11}  + \left( {Z_1^2  - Z_2 } \right)C_{00} = I_1 + I_2 +I_3+I_4+I_5+I_6+I_7+I_8
\ee
We analyze the integrals in $D_{aa,aa}$ one by one. In all the following analyses, we focus on the low-frequency regime. (High-frequency regime is relatively
straightforward, where the steepest descent method can be applied directly.) The three integrals $I_1$, $I_2$ and $I_3$ are time-independent,
\bea
I_1  &=&  - 2\int {\frac{{d\theta _1 }}
{{2\pi }}F_2^\sigma  (i\pi ,0)\left\langle \sigma  \right\rangle e^{ - 2\beta \Delta _a \cosh \theta _1 } }
\\
I_2  &=& \frac{1}
{2}\int {\int {\frac{{d\theta _1 d\theta _2 }}
{{(2\pi )^2 }}\left( {F_2^\sigma  (i\pi ,0)} \right)^2 } e^{ - \beta \Delta _a \left( {\cosh \theta _1  + \cosh \theta _2 } \right)} }
\\
I_3  &=& \int {\int {\frac{{d\theta _1 d\theta _2 }}
{{(2\pi )^2 }}F_{4s}^\sigma  (\theta _1 ,\theta _2 )\left\langle \sigma  \right\rangle } e^{ - \beta \Delta _a \left( {\cosh \theta _1  + \cosh \theta _2 } \right)} }
\eea
where
\bea
\begin{gathered}
  F_{4s}^\sigma  (\theta _1 ,\theta _2 ) = \mathop {\lim }\limits_{\varepsilon  \to 0} F_4^\sigma  \left( {\theta _1  + i\pi  + \varepsilon ,\theta _2  + i\pi  + \varepsilon ,\theta _2 ,\theta _1 } \right) \hfill \\
   = \frac{{2iF_2^\sigma  \left( {\theta _2  + i\pi ,\theta _1 } \right)\left[ {S_{aa} (\theta _1  - \theta _2 ) - S_{aa} (\theta _2  - \theta _1 )} \right]}}
{{\theta _1  - \theta _2 }} + F_{4rc}^\sigma  \left( {\theta _2  + i\pi ,\theta _1  + i\pi |\theta _2 ,\theta _1 } \right) \hfill \\
\end{gathered}
\eea
Since $\mathop {\lim }\limits_{z \to 0} \left[ {\left( {S_{aa}(z) - S_{aa}( - z)} \right)/z} \right] = 2S'_{aa}(0)$ is finite, and $F_{4rc}^{\sigma}$ is a regular function on real axis \cite{pozsgay,szecsenyi}, the whole integrand in $I_3$ is regular. As we mentioned before we will return to the discussion of these constant parts.

The integral $I_4$ is
\be
I_4  =  - \int {\int {\frac{{d\theta _1 d\theta'_1}}
{{(2\pi )^2 }}\left( {F_2^\sigma  (\theta _1  + i\pi ,\theta '_1 )} \right)^2 e^{ - 2\beta \Delta _a \cosh \theta _1 } e^{ - i x\Delta _a \left( {\sinh \theta _1  - \sinh \theta '_1 } \right)} } e^{ - it\Delta _a \left( {\cosh \theta '_1  - \cosh \theta _1 } \right)} }
\ee
$I_4$ has the same integral structure as seen in the calculation of $S_{11}$, except for a different thermal weight-factor.
So we will have a similar $\ln(\omega/T)$ divergence in the low-frequency regime as in $S_{11}$.
However, it is associated with a $e^{-2\Delta_a/T}$ factor, and thus negligible compared with $S_{11}$.

The integral $I_5$ is
\be
I_5  =  - \int {\int {\frac{{d\theta _1 d\theta _2 }}
{{(2\pi )^2 }}\left( {F_2^\sigma  (\theta _2  + i\pi ,\theta _1 )} \right)^2 e^{ - \beta \Delta _a \left( {\cosh \theta _1  + \cosh \theta _2 } \right)} e^{ - i x\Delta _a \left( {\sinh \theta _2  - \sinh \theta _1 } \right)} } e^{ - it\Delta _a \left( {\cosh \theta _1  - \cosh \theta _2 } \right)} }
\ee
Again we have a similar integrand strucdture as in $S_{11}$, and therefore the divergence in the low-frequency regime in $I_5$
will not be stronger than $\ln(\omega/T)$; the thermal factor $e^{-2\Delta_a/T}$ again makes it  negligible compared with $S_{11}$.

The integral $I_6$ is $I_6  = I_6^{(1)}  + I_6^{(2)}$, with
\bea
&&  I_6^{(1)}  = \int {\int {\frac{{d\theta _1 d\theta _2 }}
{{(2\pi )^2 }}\int {\frac{{d\theta '_1 }}
{{2\pi }}\left( {F_2^\sigma  (\theta _1  + i\pi ,\theta '_1 )} \right)^2 \left[ {\left( {1 - S(\theta '_1  - \theta _1 )S(\theta _1  - \theta _2 )} \right)\left( { - \Delta _a x\cosh \theta _1  + \Delta _a t\sinh \theta _1 } \right)} \right]} } }  \hfill \nonumber \\
&&  \;\;\;\;\;\;\;\;\;\;\;\;\;\;\;\;\;\;\;\;\;\;\;\;\;\;\;\;\;\;\;\;\;\;\;\;\;\;\;\;\;\;\;\;\;\;\;\;\;\;\;\;\;\;\;\;\;\;\;\;\;e^{ - \beta \Delta _a \left( {\cosh \theta _1  + \cosh \theta _2 } \right)} e^{ - ix\Delta _a \left( {\sinh \theta _2  - \sinh \theta '_1 } \right)} e^{ - it\Delta _a \left( {\cosh \theta '_1  - \cosh \theta _2 } \right)}  \hfill \\
&&  I_6^{(2)}  =  - \int {\int {\frac{{d\theta _1 d\theta _2 }}
{{(2\pi )^2 }}\int {\frac{{d\theta '_1 }}
{{2\pi }}\left( {F_2^\sigma  (\theta _1  + i\pi ,\theta '_1 )} \right)^2 \varphi \left( {\theta '_1  - \theta _1 } \right)S(\theta '_1  - \theta _1 )S(\theta _1  - \theta _2 )} } }  \hfill \nonumber \\
&&  \;\;\;\;\;\;\;\;\;\;\;\;\;\;\;\;\;\;\;\;\;\;\;\;\;\;\;\;\;\;\;\;\;\;\;\;\;\;\;\;\;\;\;\;\;\;\;\;\;\;\;\;\;\;\;\;\;\;\;\;\;e^{ - \beta \Delta _a \left( {\cosh \theta _1  + \cosh \theta _2 } \right)} e^{ - ix\Delta _a \left( {\sinh \theta _2  - \sinh \theta '_1 } \right)} e^{ - it\Delta _a \left( {\cosh \theta '_1  - \cosh \theta _2 } \right)}  \hfill
  \eea
where
\be
\varphi \left( {\theta '_1  - \theta _1 } \right)S(\theta '_1  - \theta _1 )S(\theta _1  - \theta _2 ) =  - \frac{i}
{{S(\theta '_1  - \theta _1 )}}\frac{{dS(\theta '_1  - \theta _1 )}}
{{d\theta '_1 }}S(\theta '_1  - \theta _1 )S(\theta _1  - \theta _2 ) =  - iS(\theta _1  - \theta _2 )\frac{{dS(\theta '_1  - \theta _1 )}}
{{d\theta '_1 }}
\ee
We will see that combining $I_6^{(1)}$ and part of $I_8$ gives zero contribution to the local dynamics. So we consider $I_6^{(2)}$.
Since the time-space oscillation factor in $I_6^{(2)}$, $e^{ - ix\Delta _a \left( {\sinh \theta _2  - \sinh \theta '_1 } \right)}
e^{ - it\Delta _a \left( {\cosh \theta '_1  - \cosh \theta _2 } \right)}$, is independent of rapidity $\theta_1$,
and the leftover integrand is regular on the real axis, we can apply
the steepest descent method for $\theta_1$ \cite{wang} with saddle point at $\theta_1 = 0$,
\be
I_6^{(2)}  \sim \sqrt {\frac{T}
{{\Delta _a }}} e^{ - \Delta _a /T} \int {\int {\frac{{d\theta _1 d\theta '_1 }}
{{(2\pi )^2 }}\left( {F_2^\sigma  (i\pi ,\theta '_1 )} \right)^2 S( - \theta _2 )\left[ {i\frac{{dS(\theta '_1 )}}
{{d\theta '_1 }}} \right]e^{ - \beta \Delta _a \cosh \theta _2 } e^{ - ix\Delta _a \left( {\sinh \theta _2  - \sinh \theta '_1 } \right)}
e^{ - it\Delta _a \left( {\cosh \theta '_1  - \cosh \theta _2 } \right)} } }
\ee
The leftover integral has similar structure as in $S_{11}$ and, in low-frequency regime,
\be
\left| {I_6^{(2)} (\omega )} \right| \sim \left| {\sqrt {\frac{T}
{{\Delta _a }}} \frac{{e^{ - 2\Delta _a /T} e^{\omega /2T} \left| {F_2^\sigma  (i\pi ,0)} \right|^2 }}
{{\Delta _a }}S(0)S'(0)\ln \frac{\omega }
{{4T}}} \right|\;\;(\omega  \ll T \ll \Delta _a )  .
\ee
Thus, its contribution to the  low-energy local dynamics is negligible compared with $S_{11}$.

The integral $I_7$ is
\be
\begin{gathered}
  I_7  = 2\int {\int {\frac{{d\theta _1 d\theta _2 }}
{{(2\pi )^2 }}P\int {\frac{{d\theta '_1 }}
{{2\pi }}F_{4ss}^\sigma  \left( {\theta _1  + i\pi ,\theta _2  + i\pi |\theta '_1 ,\theta _1 } \right)F_2^\sigma  (\theta _2  + i\pi ,\theta '_1 )} } }  \hfill \\
  \;\;\;\;\;\;\;\;\;\;\;\;\;\;\;\;\;\;\;\;\;\;\;\;\;\;\;\;\;\;\;\;\;\;\;\;\;\;\;\;\;\;\;\;\;\;\;\;\;\;\;\;e^{ - \beta \Delta _a \left( {\cosh \theta _1  + \cosh \theta _2 } \right)} e^{ - ix\Delta _a \left( {\sinh \theta _2  - \sinh \theta '_1 } \right)} e^{ - it\Delta _a \left( {\cosh \theta '_1  - \cosh \theta _2 } \right)}  \hfill \\
\end{gathered}
\ee
where
\be
F_{4ss}^\sigma  \left( {\theta _1  + i\pi ,\theta _2  + i\pi |\theta '_1 ,\theta _1 } \right) = \frac{{i\left( {S_{aa}(\theta _1  - \theta _2 ) + 1} \right)}}{{\theta _1  - \theta '_1 }}F_2^\sigma  (\theta _2  + i\pi ,\theta '_1 ) + F_{4rc}^\sigma  \left( {\theta _1  + i\pi ,\theta _2  + i\pi |\theta '_1 ,\theta _1 } \right)
\ee
Consider the part containing $F_{4rc}^\sigma  \left( {\theta _1  + i\pi ,\theta _2  + i\pi |\theta '_1 ,\theta _1 } \right)$.
Since $F_{4rc}^\sigma  \left( {\theta _1  + i\pi ,\theta _2  + i\pi |\theta '_1 ,\theta _1 } \right)$ is not singular on the real axis,
this part of the integration behaves similarly as that in $I_6^{(2)}$, making its contribution negligible in the low-frequency regime.
Consider next the part containing $\frac{{i\left( {S_{aa}(\theta _1  - \theta _2 ) + 1} \right)}}{{\theta _1  - \theta '_1 }}F_2^\sigma  (\theta _2  + i\pi ,\theta '_1 )$. The integration here is still well defined in the sense of principal-value integration over $\theta'_1$. Recall the definition of principal-value integration:
\be
P\int_L {\frac{{h(\tau )}}
{{\tau  - t}}d\tau }  = \int_L {\frac{{h(\tau ) - h(t)}}
{{\tau  - t}}d\tau }  + h(t)\ln \frac{{b - t}}
{{t - a}}
\ee
where $t$ lies on curve $L$ (not at end points). Since $\theta'_1$ integration is on real axis,
\be
P\int_R {\frac{{h(\theta '_1 )}}
{{\theta _1  - \theta '_1 }}d\theta '}  = P\int_{ - \infty }^\infty  {\frac{{h(\theta '_1 )}}
{{\theta _1  - \theta '_1 }}d\theta '_1 }  =  - \int_L {\frac{{h(\theta '_1 ) - h(\theta _1 )}}
{{\theta '_1  - \theta _1 }}d\tau }
\ee
In our case,
\be
h(\theta '_1 ) = i\left( {S_{aa}(\theta _1  - \theta _2 ) + 1} \right)\left[ {F_2^\sigma  (\theta _2  + i\pi ,\theta '_1 )} \right]^2 e^{i\Delta _a x\sinh \theta '_1 } e^{ - it\Delta _a \cosh \theta '_1 }
\ee
The function associated with $e^{-\beta \Delta_a \cosh \theta_1}$ is not singular on real axis. Thus we can apply steepest decent method for $\theta_1$ \cite{wang}, leaving us $\theta'_1$ and $\theta_2$ integrations as
\be
\begin{gathered}
  I_7  \sim 2\sqrt {\frac{T}
{{\Delta _a }}} e^{ - \beta \Delta _a } \iint_R {\frac{{d\theta _2 d\theta '_1 }}
{{(2\pi )^2 }}\frac{{i\left[ {S_{aa}( - \theta _2 ) + 1} \right]\left[ {\left( {F_2^\sigma  \left( {\theta _2  + i\pi ,\theta '_1 } \right)} \right)^2  - \left( {F_2^\sigma  \left( {\theta _2  + i\pi ,0} \right)} \right)^2 } \right]}}
{{\theta '_1 }}} \hfill \\
  \;\;\;\;\;\;\;\;\;\;\;\;\;\;\;\;\;\;\;\;\;\;\;\;\;\;\;\;\;\;\;\;\;\;\;\;\;\;\;\;\;\;\;\;\;\;\;\;\;\;\;\;\;\;\;\;\;e^{ - \beta \Delta_a \cosh \theta _2 } e^{ - i\Delta _a x\left( {\sinh \theta _2  - \sinh \theta '_1 } \right)} e^{ - it\Delta _a \left( {\cosh \theta '_1  - \cosh \theta _2 } \right)}  \hfill \\
\end{gathered}
\ee
For the leftover integration, we encounter a structure similar as in $S_{11}$. Therefore the part involving principal-value integraion
gives contribution at the order of $\frac{1}{{\Delta _a }}\sqrt {\frac{T}{{\Delta _a }}} e^{ - 2\Delta _a /T} \left| {F_2^\sigma  (i\pi ,0)}
\right|^2 \ln \frac{\omega }{{4T}}$. Combining with the other part's contribution we have
\be
I_7 (\omega ) \sim \frac{1}
{{\Delta _a }}\sqrt {\frac{T}
{{\Delta _a }}} e^{ - 2\Delta _a /T} \left| {F_2^\sigma  (i\pi ,0)} \right|^2 \ln \frac{\omega }
{{4T}}\;\;(\omega  \ll T \ll \Delta _a ).
\ee
We conclude that $I_7$'s contribution to low-energy local dynamics is negligible compared with $S_{11}$.

The integral $I_8$ is
\be
\begin{gathered}
  I_8  = \frac{1}
{4}\int {\int {\int {\int {\frac{{d\theta _1 d\theta _2 d\theta '_1 d\theta '_2 }}
{{(2\pi )^4 }}F_4^\sigma  \left( {\theta _2  + i\pi ,\theta _1  + i\pi ,\theta '_1 ,\theta '_2 } \right)F_4^\sigma  \left( {\theta _1  + i\pi ,\theta _2  + i\pi ,\theta '_2 ,\theta '_1 } \right)} } } }  \hfill \\
  \;\;\;\;\;\;\;\;\;\;\;\;\;\;\;\;\;\;\;\;\;\;\;e^{ - \beta \Delta _a (\cosh \theta _1  + \cosh \theta _2 )} e^{ - i\Delta _a x(\sinh \theta _1  + \sinh \theta _2  - \sinh \theta '_1  - \sinh \theta '_2 )} e^{ - it\Delta _a (\cosh \theta '_1  + \cosh \theta '_2  - \cosh \theta _1  - \cosh \theta _2 )}  \hfill \\
\end{gathered}
\ee
where
\bea
&&  F_4^\sigma  \left( {\theta _2  + i\pi ,\theta _1  + i\pi ,\theta '_1 ,\theta '_2 } \right)F_4^\sigma  \left( {\theta _1  + i\pi ,\theta _2  + i\pi ,\theta '_2 ,\theta '_1 } \right) \hfill \\
   &=& F_{4rc}^\sigma  \left( {\theta _2  + i\pi ,\theta _1  + i\pi |\theta '_1  + i\varepsilon ,\theta '_2  + i\varepsilon } \right)F_{4rc}^\sigma  \left( {\theta _1  + i\pi ,\theta _2  + i\pi | \theta '_2  + i\varepsilon ,\theta '_1  + i\varepsilon } \right) \hfill \label{f41} \\
&+& F_{4rc}^\sigma  \left( {\theta _2  + i\pi ,\theta _1  + i\pi |\theta '_1  + i\varepsilon ,\theta '_2  + i\varepsilon } \right)\left[ {\frac{E}
{{\theta _1  - \theta '_2  - i\varepsilon }} + \frac{F}
{{\theta _1  - \theta '_1  - i\varepsilon }} + \frac{G}
{{\theta _2  - \theta '_2  - i\varepsilon }} + \frac{H}
{{\theta _2  - \theta '_1  - i\varepsilon }}} \right] \hfill \label{f42} \\
  & +& F_{4rc}^\sigma  \left( {\theta _1  + i\pi ,\theta _2  + i\pi |\theta '_2  + i\varepsilon ,\theta '_1  + i\varepsilon } \right)\left[ {\frac{A}
{{\theta _2  - \theta '_1  - i\varepsilon }} + \frac{B}
{{\theta _2  - \theta '_2  - i\varepsilon }} + \frac{C}
{{\theta _1  - \theta '_1  - i\varepsilon }} + \frac{D}
{{\theta _1  - \theta '_2  - i\varepsilon }}} \right] \hfill  \label{f43} \\
&+& \frac{{AH}}
{{\left( {\theta _2  - \theta '_1  - i\varepsilon } \right)^2 }} + \frac{{BG}}
{{\left( {\theta _2  - \theta '_2  - i\varepsilon } \right)^2 }} + \frac{{CF}}
{{\left( {\theta _1  - \theta '_1  - i\varepsilon } \right)^2 }} + \frac{{DE}}
{{\left( {\theta _1  - \theta '_2  - i\varepsilon } \right)^2 }} + \frac{{AE + DH}}
{{\left( {\theta _2  - \theta '_1  - i\varepsilon } \right)\left( {\theta _1  - \theta '_2  - i\varepsilon } \right)}} \hfill \label{f44} \\
   &+& \frac{{AF + CH}}
{{\left( {\theta _2  - \theta '_1  - i\varepsilon } \right)\left( {\theta _1  - \theta '_1  - i\varepsilon } \right)}} + \frac{{AG + BH}}
{{\left( {\theta _2  - \theta '_1  - i\varepsilon } \right)\left( {\theta _2  - \theta '_2  - i\varepsilon } \right)}} + \frac{{BE + DG}}
{{\left( {\theta _2  - \theta '_2  - i\varepsilon } \right)\left( {\theta _1  - \theta '_2  - i\varepsilon } \right)}} \hfill \label{f45} \\
   &+& \frac{{BF + CG}}
{{\left( {\theta _2  - \theta '_2  - i\varepsilon } \right)\left( {\theta _1  - \theta '_1  - i\varepsilon } \right)}} + \frac{{CE + DF}}
{{\left( {\theta _1  - \theta '_1  - i\varepsilon } \right)\left( {\theta _1  - \theta '_2  - i\varepsilon } \right)}} \hfill. \label{f46}
\eea
with
\be
\begin{gathered}
  A = i\left[ {S\left( {\theta _2  - \theta _1 } \right) - S\left( {\theta '_1  - \theta '_2 } \right)} \right]F_2^\sigma  \left( {\theta _1  + i\pi ,\theta '_2 } \right);\;\;B = i\left[ {S\left( {\theta '_1  - \theta '_2 } \right)S\left( {\theta _2  - \theta _1 } \right) - 1} \right]F_2^\sigma  \left( {\theta _1  + i\pi ,\theta '_1 } \right); \hfill \\
  C = i\left[ {1 - S\left( {\theta _2  - \theta _1 } \right)S\left( {\theta '_1  - \theta '_2 } \right)} \right]F_2^\sigma  \left( {\theta _2  + i\pi ,\theta '_2 } \right);\;D = i\left[ {S\left( {\theta '_1  - \theta '_2 } \right) - S\left( {\theta _2  - \theta _1 } \right)} \right]F_2^\sigma  \left( {\theta _2  + i\pi ,\theta '_1 } \right); \hfill \\
  E = i\left[ {S\left( {\theta _1  - \theta _2 } \right) - S\left( {\theta '_2  - \theta '_1 } \right)} \right]F_2^\sigma  \left( {\theta _2  + i\pi ,\theta '_1 } \right);\;\;F = i\left[ {S\left( {\theta '_2  - \theta '_1 } \right)S\left( {\theta _1  - \theta _2 } \right) - 1} \right]F_2^\sigma  \left( {\theta _2  + i\pi ,\theta '_2 } \right); \hfill \\
  G = i\left[ {1 - S\left( {\theta _1  - \theta _2 } \right)S\left( {\theta '_2  - \theta '_1 } \right)} \right]F_2^\sigma  \left( {\theta _1  + i\pi ,\theta '_1 } \right);\; H = i\left[ {S\left( {\theta '_2  - \theta '_1 } \right) - S\left( {\theta _1  - \theta _2 } \right)} \right]F_2^\sigma  \left( {\theta _1  + i\pi ,\theta '_2 } \right) \hfill \\
\end{gathered}
\ee

From Eq.~(\ref{f41}) we have ($q$ and $\omega$ have been rescaled by $\Delta_a$)
\be
\begin{gathered}
  I_8^{(1)} (\omega ,q) = \frac{1}
{{\Delta _a^2 }}\int {\frac{{d\theta _1 d\theta _2 d\theta '_1 d\theta '_2 }}
{{(2\pi )^2 }}F_{4rc}^\sigma  \left( {\theta _2  + i\pi ,\theta _1  + i\pi |\theta '_1 ,\theta '_2 } \right)F_{4rc}^\sigma  \left( {\theta _1  + i\pi ,\theta _2  + i\pi ,\theta '_2 ,\theta '_1 } \right)}  \hfill \\
  \delta \left( {q + \sinh \theta _1  + \sinh \theta _2  - \sinh \theta '_1  - \sinh \theta '_2 } \right)\delta \left( {\omega  + \cosh \theta _1  + \cosh \theta _2  - \cosh \theta '_1  - \cosh \theta '_2 } \right) \hfill \\
   = \frac{1}
{{\Delta _a^2 }}\int {\frac{{d\theta _1 d\theta _2 }}
{{(2\pi )^2 }}\frac{{F_{4rc}^\sigma  \left( {\theta _2  + i\pi ,\theta _1  + i\pi |\theta '_1 ,\theta '_2 } \right)F_{4rc}^\sigma  \left( {\theta _1  + i\pi ,\theta _2  + i\pi ,\theta '_2 ,\theta '_1 } \right)e^{ - \beta \Delta _a \left( {\cosh \theta _1  + \cosh \theta _2 } \right)} }}
{{\left\{ {\left[ {\frac{{\left( {\omega  + \cosh \theta _1  + \cosh \theta _2 } \right)^2  - \left( {q + \sinh \theta _1  + \sinh \theta _2 } \right)^2 }}
{2} - 1} \right]^2  - 1} \right\}^{1/2} }}}  \hfill \\
\end{gathered}
\ee
which leads to
\be
I_8^{(1)} (\omega ) = \frac{1}
{{\Delta _a }}\int {dq\int {\frac{{d\theta _1 d\theta _2 }}
{{(2\pi )^2 }}\frac{{F_{4rc}^\sigma  \left( {\theta _2  + i\pi ,\theta _1  + i\pi |\theta '_1 ,\theta '_2 } \right)F_{4rc}^\sigma  \left( {\theta _1  + i\pi ,\theta _2  + i\pi ,\theta '_2 ,\theta '_1 } \right)e^{ - \beta \Delta _a \left( {\cosh \theta _1  + \cosh \theta _2 } \right)} }}
{{\left\{ {\left[ {\frac{{\left( {\omega  + \cosh \theta _1  + \cosh \theta _2 } \right)^2  - \left( {q + \sinh \theta _1  + \sinh \theta _2 } \right)^2 }}
{2} - 1} \right]^2  - 1} \right\}^{1/2} }}} }
\ee
where $\theta'_1$ and $\theta'_2$ are functions of $\theta_1$ and $\theta_2$. Then we can apply steepest descent method on $I_8^{(1)} (\omega )$, leading to (unlike $S_{11}$, here $\theta_1$ and $\theta_2$ are independent of $q$ and $\omega$),
\be
I_8^{(1)} (\omega ) \sim \frac{{\left( {F_{4rc}^\sigma  \left( {i\pi ,i\pi |0,0} \right)} \right)^2 }}
{{\Delta _a }}\frac{T}
{{\Delta _a }}e^{ - 2\Delta _a /T} \int {dq\frac{1}
{{\sqrt {\left[ {\left( {\frac{{\left( {\omega  + 2} \right)^2  - q^2 }}
{2}} \right) - 1} \right]^2  - 1} }}}
\ee
The allowed integration range of $q$ can be determined by
\be
\begin{gathered}
  \left[ {\left( {\frac{{\left( {\omega  + 2} \right)^2  - q^2 }}
{2}} \right) - 1} \right]^2  - 1 \geqslant 0 \Rightarrow \left( {\omega ^2  + 4\omega  - q^2 } \right)\left[ {\frac{1}
{4}\left( {\omega ^2  + 4\omega  - q^2 } \right) + 1} \right] \geqslant 0 \Rightarrow  \hfill \\
  \left( {q^2  - \omega ^2  - 4\omega } \right)\left( {q^2  - 4\omega  - \omega ^2  - 4} \right) \geqslant 0 \Rightarrow q^2  \geqslant 4 + \omega ^2  + 4\omega \;or\;q^2  \leqslant \omega ^2  + 4\omega  \hfill \\
\end{gathered}
\ee
Using evenness of the integrand as a function of $q$ (so the integral over $q$ can be shrunk to $(0, \infty)$ ) and making variable
transform $z= \omega^2 +4\omega -q^2$, we have
\be
\begin{gathered}
  I_8^{(1)} (\omega ) \sim \frac{{\left( {F_{4rc}^\sigma  \left( {i\pi ,i\pi |0,0} \right)} \right)^2 }}
{{\pi \Delta _a }}\frac{T}
{{\Delta _a }}e^{ - 2\Delta _a /T} \left( {\int_4^\infty  {dz}  + \int_{ - \left( {4\omega  + \omega ^2 } \right)}^0 {dz} } \right)\int {dq\frac{1}
{{\sqrt {\left( {z + 4\omega  + \omega ^2 } \right)\left( {z - 4} \right)z} }}}  \hfill \\
   = \frac{{\left( {F_{4rc}^\sigma  \left( {i\pi ,i\pi |0,0} \right)} \right)^2 }}
{{\pi \Delta _a }}\frac{T}
{{\Delta _a }}e^{ - 2\Delta _a /T} \left[ {\frac{{2\left( {iK\left( { - 4/a} \right) + K\left( {1 + 4/a} \right)} \right)}}
{{\sqrt a }} + K\left( { - a/4} \right)} \right]\;\;(a = \omega ^2  + 4\omega ) \hfill \\
   = \frac{{\left( {F_{4rc}^\sigma  \left( {i\pi ,i\pi |0,0} \right)} \right)^2 }}
{{\pi \Delta _a }}\frac{T}
{{\Delta _a }}e^{ - 2\Delta _a /T} \left\{ {\pi  - \frac{\pi }
{{16}}a + \frac{{9\pi }}
{{1024}}a^2  +  \cdots  \cdots } \right\}\;\;(a \ll 1) \hfill \\
   = \frac{{\left( {F_{4rc}^\sigma  \left( {i\pi ,i\pi |0,0} \right)} \right)^2 }}
{{\Delta _a }}\frac{T}
{{\Delta _a }}e^{ - 2\Delta _a /T} \left\{ {1 - \frac{1}
{4}\frac{\omega }
{{\Delta _a }} +  \cdots  \cdots } \right\}\;
\hfill \\
\end{gathered}
\ee
where $K$ is the complete elliptic integral of the first kind.Therefore, $I_8^{(1)}$ is negligible for the low-energy local dynamics compared with $S_{11}$.

For Eqs.~(\ref{f42},\ref{f43}), all terms have a similar structure, so we can just focus on one of them.
\be
\begin{gathered}
  I_8^{(2)}  = \frac{1}
{4}\int {\frac{{d\theta _1 d\theta _2 d\theta '_1 d\theta '_2 }}
{{(2\pi )^4 }}F_{4rc}^\sigma  \left( {\theta _2  + i\pi ,\theta _1  + i\pi |\theta '_1  + i\varepsilon ,\theta '_2  + i\varepsilon } \right)\frac{E}
{{\theta _1  - \theta '_2  - i\varepsilon }}}  \hfill \\
  \;\;\;\;\;\;\;\;\;\;\;\;\;\;\;\;\;e^{ - \beta \Delta _a (\cosh \theta _1  + \cosh \theta _2 )} e^{ - i\Delta _a x(\sinh \theta _1  + \sinh \theta _2  - \sinh \theta '_1  - \sinh \theta '_2 )} e^{ - it\Delta _a (\cosh \theta '_1  + \cosh \theta '_2  - \cosh \theta _1  - \cosh \theta _2 )}  \hfill \\
   = I_8^{(2),1}  + I_8^{(2),2}  \hfill \\
\end{gathered}
\ee
where
\be
\begin{gathered}
  I_8^{(2),1}  = \frac{1}
{4}P\int {\frac{{d\theta _1 d\theta _2 d\theta '_1 d\theta '_2 }}
{{(2\pi )^4 }}F_{4rc}^\sigma  \left( {\theta _2  + i\pi ,\theta _1  + i\pi |\theta '_1  + i\varepsilon ,\theta '_2  + i\varepsilon } \right)\frac{E}
{{\theta _1  - \theta '_2 }}}  \hfill \\
  \;\;\;\;\;\;\;\;\;\;\;\;\;e^{ - \beta \Delta _a (\cosh \theta _1  + \cosh \theta _2 )} e^{ - i\Delta _a x(\sinh \theta _1  + \sinh \theta _2  - \sinh \theta '_1  - \sinh \theta '_2 )} e^{ - it\Delta _a (\cosh \theta '_1  + \cosh \theta '_2  - \cosh \theta _1  - \cosh \theta _2 )}  \hfill \\
\end{gathered}
\ee
and
\be
\begin{gathered}
  I_8^{(2),2}  = \frac{1}
{4}i\pi \int {\frac{{d\theta _1 d\theta _2 d\theta '_1 d\theta '_2 }}
{{(2\pi )^4 }}F_{4rc}^\sigma  \left( {\theta _2  + i\pi ,\theta _1  + i\pi |\theta '_1  + i\varepsilon ,\theta '_2  + i\varepsilon } \right)E\delta (\theta _1  - \theta '_2 )}  \hfill \\
  \;\;\;\;\;\;\;\;\;\;\;e^{ - \beta \Delta _a (\cosh \theta _1  + \cosh \theta _2 )} e^{ - i\Delta _a x(\sinh \theta _1  + \sinh \theta _2  - \sinh \theta '_1  - \sinh \theta '_2 )} e^{ - it\Delta _a (\cosh \theta '_1  + \cosh \theta '_2  - \cosh \theta _1  - \cosh \theta _2 )}  \hfill \\
\end{gathered}
\ee
For $I_8^{(2),1}$, the principal value integral structure will be similar as that appearing in $I_7$.
Similar analysis can be applied to $I_8^{(2),1}$, leading to a non-singular contribution in the low-frequency regime
(it's a four-fold integration similar to that appearing in $I_8^{(1)}$). As for $I_8^{(2),2}$ it's easy to get
\be
I_8^{(2),2}  = \frac{1}
{4}i\pi \int {\frac{{d\theta _1 d\theta _2 d\theta '_1 }}
{{(2\pi )^4 }}F_{4rc}^\sigma  \left( {\theta _2  + i\pi ,\theta _1  + i\pi |\theta '_1 ,\theta '_2 } \right)\left. E \right|_{\theta _1  = \theta '_2 } e^{ - \beta \Delta _a (\cosh \theta _1  + \cosh \theta _2 )} e^{ - i\Delta _a x(\sinh \theta _2  - \sinh \theta '_1 )} e^{ - it\Delta _a (\cosh \theta '_1  -  \cosh \theta _2 )} }
\ee
where $\left. E \right|_{\theta _1  = \theta '_2 }  = i\left[ {S( - \theta _2 ) - S( - \theta '_1 )} \right]F_2^\sigma  \left( {\theta _2  + i\pi ,\theta '_1 } \right)$.
We encounter similar integral structure as  shown in $I_6^{(2)}$.  Thus, this part's contribution will be of the same order
as that appearing in $I_6^{(2)}$. Therefore the contribution from $I_8^{(2)}$  to the low-energy local dynamics  is negligible
compared with $S_{11}$.

For Eqs.~(\ref{f44},\ref{f45},\ref{f46}), let's first consider the parts containing terms similar to the following:
\be
\frac{{AE + DH}}
{{\left( {\theta _2  - \theta '_1  - i\varepsilon } \right)\left( {\theta _1  - \theta '_2  - i\varepsilon } \right)}}
\ee
Other five similar terms will have contribution at the same order of this one. For this one we have
\bea
  \frac{{AE + DH}}
{{\left( {\theta _2  - \theta '_1  - i\varepsilon } \right)\left( {\theta _1  - \theta '_2  - i\varepsilon } \right)}} &=& P\frac{1}
{{\theta _2  - \theta '_1 }}P\frac{1}
{{\theta _1  - \theta '_2 }}\left( {AE + DH} \right) + P\frac{1}
{{\theta _2  - \theta '_1 }}i\pi \delta (\theta _1  - \theta '_2 )\left( {AE + DH} \right) \hfill \\
 && P\frac{1}
{{\theta _1  - \theta '_2 }}i\pi \delta (\theta _2  - \theta '_1 )\left( {AE + DH} \right) - \pi ^2 \delta (\theta _2  - \theta '_1 )\delta (\theta _1  - \theta '_2 )\left( {AE + DH} \right) \hfill
\eea
For the first term we will encounter similar structure as $I_8^{(1)}$, and for the second and third terms we will encounter similar structure as $I_8^{(2)}$.
It is also easy to determine $\pi ^2 \delta (\theta _2  - \theta '_1 )\delta (\theta _1  - \theta '_2 )\left( {AE + DH} \right) = 0$. Thus,
the total contribution from the term containing $\frac{{AE + DH}}{{\left( {\theta _2  - \theta '_1  - i\varepsilon } \right)
\left( {\theta _1  - \theta '_2  - i\varepsilon } \right)}}$ is negligible.
This applies to other similar terms, in which there can exist non-vanishing terms of two multiples of delta functions.
The terms having this kind of structure will have similar integral structure as $S_{11}$,
after integrating over the two delta functions. But the thermal factor $e^{-2\Delta_a/T}$ makes this negligible.

We next discuss the last terms which have a similar structure as
\be
\frac{{AH}}
{{\left( {\theta _2  - \theta '_1  - i\varepsilon } \right)^2 }}
\ee
Such terms can formerly be handled as follows,
\be
\frac{{AH}}
{{\left( {\theta _2  - \theta '_1  - i\varepsilon } \right)^2 }} \to Integration\;by\;part \to \int {\frac{1}
{{\theta '_1  - \theta _2  + i\varepsilon }}\partial _{\theta '_1 } \left( {AH \cdots  \cdots } \right)}
\ee
Combining the contributions from four such terms with that appearing in $I_6^{(1)}$ will yield zero  contribution to the low-energy local dynamics.
Explicitly we have
\be
\begin{gathered}
  AH =  - \left[ {S(\theta _2  - \theta _1 ) - S\left( {\theta '_1  - \theta '_2 } \right)} \right]F_2^\sigma  \left( {\theta _1  + i\pi ,\theta '_2 } \right)\left[ {S(\theta '_2  - \theta '_1 ) - S(\theta _1  - \theta _2 )} \right]F_2^\sigma  \left( {\theta _1  + i\pi ,\theta '_2 } \right) \hfill \\
   = \left[ {2 - S(\theta _2  - \theta _1 )S(\theta '_2  - \theta '_1 ) - S(\theta _1  - \theta _2 )S\left( {\theta '_1  - \theta '_2 } \right)} \right]\left( {F_2^\sigma  \left( {\theta _1  + i\pi ,\theta '_2 } \right)} \right)^2  \hfill \\
\end{gathered}
\ee
$\Rightarrow$
\be
\begin{gathered}
  \frac{1}
{4}\int {\frac{{d\theta _1 d\theta _2 d\theta '_1 d\theta '_2 }}
{{(2\pi )^4 }}\frac{{AH}}
{{\left( {\theta _2  - \theta '_1  - i\varepsilon } \right)^2 }}K_{tx}^{(\beta )} \left( {\theta _1 \theta _2 |\theta '_1 \theta '_2 } \right)}  \hfill \\
   = \frac{1}
{4}\int {\frac{{d\theta _1 d\theta _2 d\theta '_2 }}
{{(2\pi )^4 }}\left. {\frac{{AHK_{tx}^{(\beta )} \left( {\theta _1 \theta _2 |\theta '_1 \theta '_2 } \right)}}
{{\theta _2  - \theta '_1  - i\varepsilon }}} \right|_{\theta '_1  =  - \infty }^{\theta '_1  = \infty } }
  + \frac{1}
{4}\int {\frac{{d\theta _1 d\theta _2 d\theta '_1 d\theta '_2 }}
{{(2\pi )^4 }}\frac{1}
{{\theta '_1  - \theta _2  + i\varepsilon }}\left[ {K_{tx}^{(\beta )} \partial _{\theta '_1 } \left( {AH} \right) + AH\left( {\partial _{\theta '_1 } K_{tx}^{(\beta )} } \right)} \right]}  \hfill \\
   = \frac{1}
{4}\int {\frac{{d\theta _1 d\theta _2 d\theta '_1 d\theta '_2 }}
{{(2\pi )^4 }}\left[ {P\frac{1}
{{\theta '_1  - \theta _2 }} - i\pi \delta \left( {\theta '_1  - \theta _2 } \right)} \right]\left[ {K_{tx}^{(\beta )} \partial _{\theta '_1 } \left( {AH} \right) + AH\left( {\partial _{\theta '_1 } K_{tx}^{(\beta )} } \right)} \right]}  \hfill \\
\end{gathered}
\ee
where
\be
K_{tx}^{(\beta )} \left( {\theta _1 \theta _2 |\theta '_1 \theta '_2 } \right) = e^{ - \beta \Delta _a (\cosh \theta _1  + \cosh \theta _2 )} e^{ - i\Delta _a x(\sinh \theta _1  + \sinh \theta _2  - \sinh \theta '_1  - \sinh \theta '_2 )} e^{ - it\Delta _a (\cosh \theta '_1  + \cosh \theta '_2  - \cosh \theta _1  - \cosh \theta _2 )}
\ee
Let's focus on the following integral (all other integrals will have similar features as before),
\be
\frac{1}
{4}\int {\frac{{d\theta _1 d\theta _2 d\theta '_1 d\theta '_2 }}
{{(2\pi )^4 }}\left[ { - i\pi \delta \left( {\theta '_1  - \theta _2 } \right)} \right]\left[ {AH\left( {\partial _{\theta '_1 } K_{tx}^{(\beta )} } \right)} \right]}
\ee
where
\be
\begin{gathered}
  \left[ { - i\pi \delta \left( {\theta '_1  - \theta _2 } \right)} \right]\left[ {AH\left( {\partial _{\theta '_1 } K_{tx}^{(\beta )} } \right)} \right] = \pi \delta \left( {\theta '_1  - \theta _2 } \right)\left[ {2 - S(\theta _2  - \theta _1 )S(\theta '_2  - \theta _2 ) - S(\theta _1  - \theta _2 )S\left( {\theta _2  - \theta '_2 } \right)} \right] \hfill \\
  \;\;\;\;\;\;\;\;\;\;\;\;\;\;\;\;\;\;\;\;\;\;\;\;\;\;\;\;\;\;\;\;\;\;\;\;\;\;\;\;\;\;\left( {F_2^\sigma  \left( {\theta _1  + i\pi ,\theta '_2 } \right)} \right)^2 \left( {x\Delta _a \cosh \theta _2  - t\Delta _a \sinh \theta _2 } \right)K_{tx}^{(\beta )} \left( {\theta _1 \theta _2 |\theta _2 \theta '_2 } \right) \hfill \\
\end{gathered}
\ee
Substituting the above results back into the integral, and after finishing the integration over the delta function we can re-label
the integral variables as follows
\be
\theta _1  \leftrightarrow \theta _2 \;and\;\theta '_2  \to \theta '_1
\ee
we get
\be
\begin{gathered}
  \frac{1}
{4}\int {\frac{{d\theta _1 d\theta _2 d\theta '_1 d\theta '_2 }}
{{(2\pi )^4 }}\left[ { - i\pi \delta \left( {\theta '_1  - \theta _2 } \right)} \right]\left[ {AH\left( {\partial _{\theta '_1 } K_{tx}^{(\beta )} } \right)} \right]}  =  \hfill \\
  \frac{1}
{8}\int {\frac{{d\theta _1 d\theta _2 d\theta '_1 }}
{{(2\pi )^3 }}\left[ {2 - S(\theta _1  - \theta _2 )S(\theta '_1  - \theta _1 ) - S(\theta _2  - \theta _1 )S(\theta _1  - \theta '_1 )} \right]}  \hfill \\
  \;\;\;\;\;\;\;\;\;\;\;\;\;\;\;\;\;\;\;\;\;\;\;\;\;\;\;\;\;\;\;\left( {F_2^\sigma  \left( {\theta _1  + i\pi ,\theta '_2 } \right)} \right)^2 \left( {x\Delta _a \cosh \theta _1  - t\Delta _a \sinh \theta _1 } \right)K_{tx}^{(\beta )} \left( {\theta _1 \theta _2 |\theta _1 \theta '_1 } \right) \hfill \\
\end{gathered}
\ee
For the other three similar terms, one can get a similar integral as above for the part we are interested in.
These parts can be combined with that appearing in $I_6^{(1)}$ and yield
\be
\begin{gathered}
  I_c (x,t) \equiv I_8^{part}  + I_6^{(1)}  = \frac{1}
{2}\int {\frac{{d\theta _1 d\theta _2 d\theta '_1 }}
{{(2\pi )^3 }}\left[ {S(\theta _1  - \theta _2 )S(\theta '_1  - \theta _1 ) - S(\theta _2  - \theta _1 )S(\theta _1  - \theta '_1 )} \right]}  \hfill \\
  \;\;\;\;\;\;\;\;\;\;\;\;\;\;\;\;\;\;\;\;\;\;\;\;\;\;\;\left( {F_2^\sigma  \left( {\theta _1  + i\pi ,\theta '_2 } \right)} \right)^2 \left( {x\Delta _a \cosh \theta _1  - t\Delta _a \sinh \theta _1 } \right)K_{tx}^{(\beta )} \left( {\theta _1 \theta _2 |\theta _1 \theta '_1 } \right) \hfill \\
\end{gathered}
\ee
$\Rightarrow$
\be
I_c (\omega ) = \frac{1}
{2}\int {\frac{{d\theta _1 d\theta _2 d\theta '_1 }}
{{(2\pi )^3 }}u(\theta _1 ,\theta _2 ,\theta '_1 ,\omega )}
\ee
with
\be
\begin{gathered}
  u(\theta _1 ,\theta _2 ,\theta '_1,\omega ) = \left( {F_2^\sigma  \left( {\theta _1  + i\pi ,\theta '_2 } \right)} \right)^2 e^{ - \beta \Delta _a (\cosh \theta _1  + \cosh \theta _2 )} e^{ - it\Delta _a \left( {\cosh \theta '_1  - \cosh \theta _2 } \right)}  \cdot  \hfill \\
  \;\;\;\;\;\;\;\;\;\;\;\;\;\;\;\;\;\;\;\;\;\;\; \cdot \left[ {S(\theta _1  - \theta _2 )S(\theta '_1  - \theta _1 ) - S(\theta _2  - \theta _1 )S(\theta _1  - \theta '_1 )} \right]\frac{{2\pi \delta \left[ {\omega  - \Delta _a \left( {\cosh \theta '_1  - \cosh \theta _2 } \right)} \right]}}
{{\omega  - \Delta _a \left( {\cosh \theta '_1  - \cosh \theta _2 } \right)}} \hfill \\
\end{gathered}
\ee
Because
\be
u( - \theta _1 , - \theta _2 , - \theta '_1 ,\omega ) =  - u(\theta _1 ,\theta _2 ,\theta '_1 ,\omega ),
\ee
we have $I_c (\omega ) = 0$.

Combining all of the above, we conclude that (except for the time-independent parts in $I_8$, see below)
there are no singularities in the frequency dependence that are stronger than that
of $S_{11}$, and the thermal factor $e^{-2\Delta_a/T}$ makes $S_{22}$ to be negligible compared to $S_{11}$.

\section{D\MakeLowercase{isconnected} C\MakeLowercase{ontributions}
\MakeLowercase{up} \MakeLowercase{to} $D_{22}$}
At $x \to \infty$  we expect the following cluster property,
\be
\left\langle {\sigma (x,t)\sigma (0,0)} \right\rangle _T  \sim \left\langle {\sigma (0,0)} \right\rangle _T^2 \label{cluster}
\ee
Applying the Leclair-Mussardo formula \cite{pozsgay2} for the single-point function $\left\langle {\sigma (0,0)} \right\rangle _T$
in Eq.~(\ref{cluster}), we can get the part which contributes time-independent pieces in the two-point correlation function
$\left\langle {\sigma (x,t)\sigma (0,0)} \right\rangle _T$. Indeed in the $E_8$ model, up to $e^{-3\Delta_i/T}\;(i=a,b,c)$,
the time independent parts up to $D_{22}$ can be summed over to $\left\langle \sigma  \right\rangle _{T,i}^2 + O(e^{-3\Delta_i/T})  $ with \cite{pozsgay,szecsenyi}
\be
\begin{gathered}
  \left\langle \sigma  \right\rangle _{T,i}  = \left\langle \sigma  \right\rangle _0  + \int {\frac{{d\theta _1 }}
{{2\pi }}F_2^\sigma  (i\pi ,0)e^{ - \beta \Delta _i \cosh \theta _1 } }  - \int {\frac{{d\theta _1 }}
{{2\pi }}F_2^\sigma  (i\pi ,0)e^{ - 2\beta \Delta _i \cosh \theta _1 } } \; \; \; \; \; \; \; \; \; \; \; \; \; \\
   + \frac{1}
{2}\int {\int {\frac{{d\theta _1 d\theta _2 }}
{{\left( {2\pi } \right)^2 }}F_{4s}^\sigma  } (\theta _1 ,\theta _2 )e^{ - \beta \Delta _i (\cosh \theta _1  + \cosh \theta _2 )} }  + O(e^{-3\Delta_i/T})\;(i=a,b,c)  \\
\end{gathered}
\ee
It's easy to see that the expressions above for
$\left\langle \sigma  \right\rangle _{T,i}$ correspond
term-by-term
to Leclair-Mussardo formula \cite{pozsgay2}
We thus expect that, when summing over to infinite terms of the expansion series, the contribution from all of these space-time independent
terms will sum over to $\left\langle {\sigma (0,0)} \right\rangle _T^2$.
In other words, none of the time-independent terms in the two-point correlation function will appear in the two-point connected correlation function.

\end{document}